\documentclass[twocolumn]{aastex62}
\usepackage{times}
\usepackage{amsmath}
\usepackage{textcomp}
\usepackage{amssymb}
\usepackage[flushleft]{threeparttable}
\usepackage{boldline}
\usepackage{tabularx}
\usepackage{lineno}
\usepackage{capt-of}
\usepackage{breakcites}
\usepackage{multirow}
\usepackage[breaklinks=true]{hyperref}
\usepackage[hyphenbreaks]{breakurl}
\usepackage{color, colortbl}
\usepackage{array}
\newcolumntype{+}{>{\global\let\currentrowstyle\relax}}
\newcolumntype{-}{>{\currentrowstyle}}
\newcommand{\rowstyle}[1]{\gdef\currentrowstyle{#1}%
  #1\ignorespaces
}
\newcommand{\Msun}{\ensuremath{M_{\odot}} }
\newcommand{\lum}{erg\,s$^{-1} $}
\newcommand{\fermi}{{\it Fermi}}
\newcommand{\nustar}{{\it NuSTAR}}
\newcommand{\xmm}{{\it XMM-Newton }}
\newcommand{\swift}{{\it Swift}}

\newcommand{\ergflux}{\mbox{${\rm \, erg \,\, cm^{-2} \, s^{-1}}$}}
\newcommand{\gm}{$\gamma$}

\shorttitle{MeV \nustar~Blazars}
\shortauthors{Marcotulli et al.}

\begin{document}

\title{\nustar~perspective on high-redshift MeV blazars}
\author{L. Marcotulli} 
\affil{Department of Physics and Astronomy, Clemson University, Kinard Lab of Physics, Clemson, SC 29634-0978, USA}
\author{V. Paliya}
\affil{Deutsches Elektronen Synchrotron DESY, Platanenallee 6, 15738 Zeuthen, Germany}
\author{M. Ajello} 
\affil{Department of Physics and Astronomy, Clemson University, Kinard Lab of Physics, Clemson, SC 29634-0978, USA}
\author{A. Kaur} 
\affil{Department of Astronomy and Astrophysics, 525 Davey Lab, Pennsylvania State University, University Park, PA16802, USA}
\author{S. Marchesi} 
\affil{Department of Physics and Astronomy, Clemson University, Kinard Lab of Physics, Clemson, SC 29634-0978, USA}
\affil{INAF-Osservatorio Astronomico di Bologna, Via Ranzani 1, 40127 Bologna, Italy}
\author{M. Rajagopal} 
\affil{Department of Physics and Astronomy, Clemson University, Kinard Lab of Physics, Clemson, SC 29634-0978, USA}
\author{D. Hartmann} 
\affil{Department of Physics and Astronomy, Clemson University, Kinard Lab of Physics, Clemson, SC 29634-0978, USA}
\author{D. Gasparrini} 
\affil{Space  Science  Data  Center  -  Agenzia  Spaziale  Italiana,  Via  del  Politecnico,  snc,  I-00133,  Roma,  Italy}
\affil{Istituto  Nazionale  di  Fisica  Nucleare,  Sezione  di  Perugia,  I-06123  Perugia,  Italy}
\author{R. Ojha} 
\affil{National Aeronautics and Space Administration, Goddard Space Flight Center, Greenbelt, MD 20771, USA}
\affil{CRESST II/ University of Maryland, Baltimore County, 1000 Hilltop Cir, Baltimore, MD 21250, USA}
\author{G. Madejski} 
\affil{Kavli Institute for Particle Astrophysics and Cosmology, SLAC National Accelerator Laboratory, Menlo Park, CA 94025, USA}

\email{lmarcot@g.clemson.edu, vaidehi.s.paliya@gmail.com}
%\linenumbers

\begin{abstract}
With bolometric luminosities exceeding $10^{48}$ erg s$^{-1}$, powerful jets and supermassive black holes at their center, MeV blazars are some of the most extreme sources in the Universe. Recently, the \fermi-{Large Area Telescope} detected five new \gm-ray emitting MeV blazars beyond redshift $z=3.1$. With the goal of precisely characterizing the jet properties of these extreme sources, we started a multiwavelength campaign to follow them up with joint \nustar, \swift~and SARA observations. We observe six high-redshift quasars, four of them belonging to the new \gm-ray emitting MeV blazars. Thorough X-ray analysis reveals spectral flattening at soft X-ray for three of these objects. The source NVSS J151002$+$570243 also shows a peculiar re-hardening of the X-ray spectrum at energies $E>6\,\rm keV$. Adopting a one-zone leptonic emission model, this combination of hard X-rays and \gm-rays enables us to determine the location of the Inverse Compton peak and to accurately constrain the jet characteristics. In the context of the jet-accretion disk connection, we find that all six sources have jet powers exceeding accretion disk luminosity, seemingly validating this positive correlation even beyond $z>3$. Our six sources are found to have $10^9\Msun$ black holes, further raising the space density of supermassive black holes in the redshift bin $z=[3,4]$.   
\end{abstract}


\section{Introduction}\label{sec:intro}

The hard X-ray band has been crucial to studying some of the most powerful blazars \citep[see e.g.][]{Tavecchio_2000,2004A&A...413..489M,2004A&A...422..103M,2005A&A...433.1163D,2006A&A...448..861M}. More recently, the outstanding sensitivity of the Nuclear Spectroscopic Telescope Array (\nustar, 3-79\,keV, \citealp{2013ApJ...770..103H})
has enabled us to find and study some of the most distant and luminous ones \citep[e.g.,][]{2013ApJ...777..147S,2015ApJ...807..167T, 2016ApJ...826...76A, 2016ApJ...825...74P,2016MNRAS.462.1542S,2017ApJ...839...96M}.
Harboring highly relativistic jets pointed closely at the observer ($\theta_V < 1/\Gamma$, $\theta_V$ being the viewing angle and $\Gamma$ the bulk Lorentz factor, $\Gamma \sim 10-15$, \citealp{1995PASP..107..803U}), this subclass of the Active Galactic Nuclei (AGNs) is home to some of the most energetic particle acceleration and radiation processes known in astrophysics. 
The boost in flux ascribed to relativistic beaming, arising from the peculiar orientation of the jets, renders them visible at redshifts well beyond $z=2$ \citep[the farthest blazar detected so far is at $z=5.47$,][]{2004ApJ...610L...9R}, making them extraordinary beacons to explore the early Universe.
Their typical double-hump spectral energy distribution (SED) spans the whole electromagnetic spectrum and is shaped 
by the non-thermal processes occurring in the jets. 
Relativistic electrons, spiraling along the magnetic field lines, undergo both synchrotron and inverse Compton (IC) process. The first produces a peak located between infrared to X-ray frequency. The second instead results in a peak located between X- and \gm-ray energies. If the electrons interact with a source of low-energy photons within the jet, this is referred to as Synchrotron Self Compton (SSC, \citealp[e.g.][]{1989ApJ...340..181G}), whereas if the photons are external to the jet (i.e.\ the accretion disk, the torus, and/or the broad line region, BLR), it is referred to as External Compton process (EC, \citealp[e.g.][]{1994ApJ...421..153S}).
Based on their optical spectra, blazars are usually classified either as BL Lacertae objects (BL Lacs) or flat spectrum radio quasars (FSRQs), the first showing weak or no emission lines, the second showing broad ($\rm EW>5\AA$) ones. 
Following \citealp{2010ApJ...716...30A}, these sources can also be classified according to the position of the synchrotron peak ($\nu^{\rm S}_{\rm peak}$), with low-, intermediate- and high-synchrotron peak (LSP, ISP, HSP) blazars  having, respectively, $\nu^{\rm S}_{\rm peak}<10^{14}$\,Hz, $10^{14}<\nu^{\rm S}_{\rm peak}<10^{15}$\,Hz and $\nu^{\rm S}_{\rm peak}>10^{15}$\,Hz. 
FSRQs usually belong to the LSP class and at the high-luminosity end of such sub-class are the so-called `MeV blazars', whose high-energy peak falls in (or close to) the MeV band.
With bolometric luminosities exceeding $10^{48}$ erg s$^{-1}$, these are among the most powerful objects in the Universe. In fact, they host powerful relativistic jets \citep[][]{2014Natur.515..376G}, are usually found at high-redshift \citep[$z>2$, e.g.,][]{2009ApJ...699..603A,2010MNRAS.405..387G,2017ApJ...839...96M,2017ApJ...837L...5A} and typically host billion solar mass black-holes \citep[e.g.,][]{2010MNRAS.405..387G,2017ApJ...851...33P}.

In accordance with the {\it blazar sequence} \citep{1998MNRAS.301..451G}, as both source luminosity and distance increase, the two SED peaks drift towards lower energies and the Compton Dominance (CD, ratio between IC and synchrotron peak luminosities) increases (CD$>$1). As a consequence, MeV blazars are bright in the hard X-rays (above 10\,keV) and therefore represent perfect targets for \nustar. Moreover, high-energy emission dominates their energetics \citep[see][for a recent review]{2016ARA&A..54..725M}, hence a good coverage of the IC hump is required in order to obtain a precise determination of the peak position. This, in turn, allows us to constrain jet parameters such as viewing angle, power, bulk Lorentz factor, as well as the shape of underlying electron distribution \citep[see e.g.\,][]{2008MNRAS.385..283C, 2009MNRAS.397..985G, Gao_2011, 2015AJ....150....8C, 2016ApJ...830...94F, 2019A&A...627A..72G, 2019arXiv190701082P}, keys to untangle the physics of these cosmic monsters. Undoubtedly, \nustar~ is the best instrument to sample the rising part of their high-energy spectrum, also enabling us to detect possible hard X-ray peculiarities of these sources (e.g.\ flattening of the spectrum, \citealp[see][]{2016ApJ...825...74P}, and variability, \citealp[e.g.][]{2016MNRAS.462.1542S}). 
Lacking an all sky MeV instrument (e.g.\ AMEGO, \citealp{1748-0221-12-11-C11024}), complementary to hard X-ray observations are the \gm-ray ones obtained with the \fermi-Large Area Telescope \citep[\fermi-LAT;][]{2009ApJ...697.1071A}. Encompassing a range of energies from 20 MeV to $>$300\,GeV, it allows the sampling of the falling part of the IC hump \citep[e.g.][]{2015ApJ...810...14A} and, together with \nustar, it is fundamental to place tight constraint on the jet power.

On another important note, the SED peaks' shift also reveals the underlying thermal emission from the accretion disk at optical-UV energies \citep[e.g.,][]{2010MNRAS.405..387G}, which can be modeled to extract the black-hole mass and the accretion disk luminosity \citep[][]{1973A&A....24..337S}. This fact renders MeV blazars excellent probes
for addressing open issues such as the accretion disk-jet connection in jetted quasars \citep[e.g.,][]{2017ApJ...851...33P}, as well as the understanding of black hole growth and evolution in the early stages of the Universe \citep[][]{2011MNRAS.416..216V}. Indeed, since the detection of a single  source with jets aligned to our line of sight implies the existence of 2$\Gamma^2$ similar sources, at the same $z$, with same black hole masses but misaligned jets, finding more high-redshift blazars allows us to probe their intrinsic population and constrain the space density of supermassive black-holes \citep[e.g.,][]{2014arXiv1410.0364S}.

Even though blazars are the most numerous extragalactic sources in \fermi-LAT catalogs \citep[see ][]{2010ApJS..188..405A, 2012ApJS..199...31N, 2015ApJS..218...23A}, few are detected at $z>3$, probably since their peak high-energy emission lies below the \fermi-LAT band. Using the improved Pass 8 dataset,
taking advantage of the higher detection threshold of the \fermi-LAT, five new \gm-ray emitting MeV blazars have been found to lie at $z>3$, the farthest located at $z=4.31$ \citep{2017ApJ...837L...5A}.
Since, on average, high-redshift sources lack good quality multi-frequency observations, we started a multiwavelength campaign in order to characterize these newly found \gm-ray emitting blazars. In particular, we wanted to obtain good X-ray spectral coverage, and, importantly, sensitive hard X-ray data, in order to derive their X-ray properties and tightly constrain the jet emission.

In this paper, we present the broadband follow up of six sources: NVSS J013126$-$100931 ($z=3.51$, hereafter J013126$-$100931), NVSS J020346$+$113445 ($z=3.63$; hereafter J020346$+$113445), NVSS J064632$+$445116 ($z=3.39$; hereafter J064632$+$445116), NVSS J135406$-$020603 ($z=3.71$, hereafter J135406$-$020603), NVSS J151002$+$57\-0243 ($z=4.31$; hereafter J151002$+$570243) and NVSS J212912$-$153841 ($z=3.26$, hereafter J212912$-$153841). Four of these objects belong to the newly detected \gm-ray blazars, while two are candidate \gm-ray sources lying just below the LAT detection threshold.
In this work, we publish their first $>$10\,keV detection with \nustar\footnote{These sources were observed as part of our \nustar~proposals 2011, 4301; P.I.: Ajello M., Marcotulli L.}. Since
blazars are extremely variable sources in all energy ranges \citep[e.g.][]{2010ApJ...722..520A,2013ApJS..207...28S,2016A&A...591A..21M, 2017ApJ...844...32P}, to accurately constrain the sources' SED and to reliably measure the X-ray spectral parameters, we also obtain quasi-simultaneous observations from a variety of facilities.
The soft X-ray and UV-optical bands are covered by the Neil Gehrels \swift~Observatory X-ray Telescope \citep[\swift-XRT;][]{2005SSRv..120..165B} and Ultraviolet and Optical Telescope \citep[\swift-UVOT;][]{2005SSRv..120...95R}.
With the aim of studying possible long-term variability properties of these sources in the soft X-ray band,
we check if multiple \swift-XRT, as well as {\it Chandra} \citep{2000SPIE.4012....2W} and {\it XMM-Newton} \citep{2001A&A...365L...1J} observations are available. All sources but J151002$+$57-0243 and J135406$-$020603 have multiple \swift-XRT observations, while {\it Chandra} and {\it XMM-Newton} data are found for J151002$+$57-0243 and J212912$-$153841.
We complement UVOT observations with quasi-simultaneous ones from the Southeastern Association for Research in Astronomy \citep[SARA,][]{2017PASP..129a5002K} facilities, in order to detect the disk emission peak.
The \gm-ray spectra of the sources are taken from \citealp{2017ApJ...837L...5A}. The data reduction pipelines are given in Section~\ref{sec:data}. In order to derive the important physical parameters of these objects, we perform the SED modeling (described in Section~\ref{sec:model}). Section~\ref{sec:result} details the results of the analysis, which are then discussed in Section~\ref{sec:disc}. 
Throughout the paper, we use cosmological 
parameters $H_0=67.8$~km~s$^{-1}$~Mpc$^{-1}$, $\Omega_m = 0.308$, and $\Omega_\Lambda = 0.692$ \citep{2016A&A...594A..13P}.

\section{Observations and data analysis}\label{sec:data}
\subsection{\fermi-LAT}
\citealp{2017ApJ...837L...5A} recently discovered five new \gm-ray emitting high-redshift blazars. In our work, we study four of these \fermi-LAT detected sources, namely J064632$+$445116, J135406$-$020603, J151002$+$570243 and J212912$-$153841, extracting their \gm-ray data directly from \citealp{2017ApJ...837L...5A}. Two additional sources, J013126$-$100931 and J020346$+$113445, were also observed. They are candidate \gm-ray emitting blazars that fall below the LAT detection threshold
(test statistic, $TS<25$, \citealp{1996ApJ...461..396M}). 
For these sources we employ the 10-years LAT sensitivity curve\footnote{\burl{http://www.slac.stanford.edu/exp/glast/groups/canda/lat_Performance.htm}}.

\subsection{\nustar}
\nustar~observed the six blazars for net exposures of $\sim30-50\,\rm ks$ during the dates reported in Table~\ref{tab:xrayobs}.
We process the data for both instrument Focal Plane Modules (FPMA and FPMB; \citealp{2013ApJ...770..103H}) using the \nustar~Data Analysis Software (\texttt{NUSTARDAS}) v1.8.0.
The event files are calibrated using the task \texttt{nupipeline}, with the response file taken from the latest Calibration Database (CALDB, v.20180419).  The generation of source 
and background spectra, ancillary and response matrix files, is achieved using the \texttt{nuproducts} script. 
For source regions, we select circles with radii of 20\arcsec~centered on the 
targets; the background events are extracted from circles of 30\arcsec~radii located in a source-free region on the same frame.
The excellent sensitivity of \nustar~provides us with a good signal-to-noise detection of all sources. Therefore, if the source has also good soft X-ray photon statistics (see Section~\ref{sec:swift}-\ref{sec:chan}), \nustar~spectra are rebinned to have at least 15 counts per bin.Visual inspection of the light-curves shows no detected variability for all six sources.
\begin{table}[t!]
 \caption{Table of simultaneous \nustar~and \swift-XRT Observations}\label{tab:xrayobs}
\hspace{-1.5cm}
 \resizebox{0.57\textwidth}{!}{
 \begin{tabular}{ c c c c c c } 
 \hline
 \hline
 \multicolumn{6}{c}{\nustar~$+$ \swift-XRT} \\
 \hline
  Name  & $N_{\rm H}$ \tablenotemark{a} & z & Exp. time (ks) & Exp. time (ks) & Date \\  
        & & & (\nustar) & (\swift-XRT)& \\
  \hline
  J013126$-$100931 & $3.40$ & 3.51 & 30 & 3.4 & 2018-07-21\\
  J020346$+$113445 & $5.96$ & 3.63 & 33 & 1.7 & 2016-11-08\\
  J064632$+$445116 & $10.7$ & 3.39 & 32 & 1.8 & 2016-11-28\\
  J135406$-$020603 & $3.35$ & 3.71 & 32 & 1.9 & 2017-05-05\\
  J151002$+$570243 & $1.57$ & 4.31 & 53 & 2.0 & 2017-04-30\\
  J212912$-$153841 & $4.85$ & 3.26 & 37 & 2.6 & 2018-09-26\\
  \hline
\multicolumn{6}{l}{
  \begin{minipage}{0.5\textwidth}
 \tablenotetext{} {\bf Notes:}
 \tablenotetext{a}{Galactic column density in units of $10^{20}$ {\mbox{${\rm \, cm^{-2}}$}} \citep{2005A&A...440..775K}.}
\end{minipage}
}\\
 \end{tabular}
}
\end{table}
\subsection{\swift}\label{sec:swift}
\nustar~observations are supplemented by quasi-simulta\-neous \swift~pointings, each of $\sim2-3\,\rm ks$ (see Table~\ref{tab:xrayobs}).
The \swift-XRT observations are executed in photon counting mode. The data are analysed with the
\texttt{XRTDAS} software package included in HEASoft (v.3.4.1) in combination with the latest calibration database CALDB (v.20180710). The event files are calibrated and cleaned using the standard pipeline task \texttt{xrtpipeline}. 
With the tool \texttt{xselect}, we extract source and background
regions: the source region was selected as a circle of 
10\arcsec~radius, whereas an annular region of inner and outer radii of 20\arcsec~and 50\arcsec, respectively, are used for the background.
Both regions are centered at the source position.
The ancillary response files are generated using \texttt{xrtmkarf}, while the response matrix files are already included in the CALDB.
In order to study possible long-term soft X-ray variability, if more than one exposure is present in the archival data, we combine all event files and produce light-curves using \texttt{xselect}. In case no significant variability is found, 
all event files are considered to extract sources' spectra and all images are added (using the summation task \texttt{sum\_ima}) in order to increase the signal of the detection\footnote{This was the case for J013126$-$100931, J020346$+$113445, J064632$+$445116 and J212912$-$153841. In particular, J020346$+$113445 is very faint, having $\sim5\times10^{-3}$ cts s$^{-1}$.}. The sources' spectra with high signal-to-noise are rebinned to have at least 15 counts per bin, while for faint sources we use one count per bin.  

For \swift-UVOT, we follow the standard pipeline procedure \cite{2008MNRAS.383..627P} to extract the final products. We use the tasks \texttt{uvotimsum}
and \texttt{uvotsource} to extract the sources' magnitude in all filters. We correct the magnitudes for Galactic extinction following the recommendations in \citet{2008AIPC.1065...81R} and subsequently derive the fluxes using the zero points listed in \citet{2011AIPC.1358..373B}. For the fainter sources (J135406$-$020603 and J151002$+$570243)  we could only derive upper limits.
\begin{table}[t!]
 \caption{Table of SARA magnitudes}\label{tab:sara}
\hspace{-1.6cm}
\resizebox{0.58\textwidth}{!}{
 \begin{tabular}{ c c c c c c } 
 \hline
 \hline
  \multicolumn{6}{c}{SARA} \\
  \hline
  Name & {UT Date\tablenotemark{a}} & \multicolumn{4}{c}{AB Magnitude\tablenotemark{b}} \\
       &  & $B$ & $V$ & $R$ & $I$ \\

  \hline
   J064632$+$445116  & 2016-12-29  &    ...          & $18.10\pm 0.02$ &  $17.92\pm 0.02$  &  $19.48 \pm 0.03$\\
   J135406$-$020603  & 2017-05-06  & $ 20.33\pm 0.08$ & $ 19.46\pm 0.05$ & $ 19.24\pm 0.03$ & $ 19.32\pm 0.26$ \\
   J151002$+$570243  & 2017-04-30  & $ 23.41\pm 0.47$ & $ 20.72\pm 0.06$ & $ 20.16\pm 0.04$ & $ 19.17\pm 0.05$ \\
  \hline
	             &             &     {\it g} & {\it r} & {\it i} & {\it z} \\
  \hline 
   J013126$-$100931  & 2018-11-17  & $21.43\pm 0.22$ & $21.54\pm0.46$ & $19.41\pm0.08$ & $18.88\pm0.07$\\
   J212912$-$153841  & 2018-11-17  & $18.26\pm0.05$ & $16.98\pm0.01$& $16.84\pm0.01$ & $17.26\pm0.02$\\ 
\multicolumn{6}{l}{%
  \begin{minipage}{0.5\textwidth}
 \tablenotetext{} {\bf Notes:}
 \tablenotetext{a}{Exposure start time.} 
 \tablenotetext{b}{Corrected for Galactic reddening.}
\end{minipage}%
}\\
 \end{tabular}}
\end{table}

\subsection{{\it Chandra} and {\it XMM-Newton}}\label{sec:chan}
In order to study any possible long-term soft X-ray variability properties of the sources, 
we looked for archival observations with {\it Chandra} \citep{2000SPIE.4012....2W} and {\it XMM-Newton} \citep{2001A&A...365L...1J}, which we find only for J151002$+$570243 and J212912$-$153841. 

The {\it Chandra} data are reduced using the \texttt{CIAO 4.9} software \citep{2006SPIE.6270E..1VF} and the {\it Chandra} CALDB (v.4.7.8), following standard procedures (\texttt{chandra\_repro}). The source and background spectra are extracted using, respectively, a circular region of radius 2\arcsec~and an annulus of inner radius 7\arcsec~and outer radius of 15\arcsec, after a visual inspection to avoid contamination from nearby sources. The spectra are obtained using the task \texttt{specextract}.

For {\it XMM} we use the {\it XMM-Newton} Science Analysis Software (SAS) v16.0.0. The instrument has onboard the European Photon Imaging Camera (EPIC), which consists of two CCD arrays, MOS and pn. We use the EPIC/pn, EPIC/MOS1 and EPIC/MOS2 
camera observations for our purposes. 
Standard procedures are followed to reduce the data: event files are generated using \texttt{emproc}/\texttt{epproc} (for MOS and PN, respectively); after subtracting bad pixels and bad events from the field, source and background spectra are obtained using \texttt{evselect}. The source and background regions are selected from a circle of radius 10\arcsec~and an annulus of inner radius 20\arcsec~and outer radius 40\arcsec, both centered on the source of interest.  
Both {\it Chandra} and {\it XMM} spectra are rebinned to have at least 15 counts per bin. 

\subsection{SARA}
We carry out quasi-simultaneous optical observations for five objects\footnote{Due to bad weather, J013126-100931, J064632$+$445116 and J212912-153841 were observed within few days/months to \nustar~pointing, while J020346$+$113445 could not be observed due to instrumental issues.} with the Southern Association for Research in Astronomy (SARA, \citealp{2017PASP..129a5002K}) consortium's 1.0 m, 0.96 m and 0.6 m telescopes respectively located at la Roque de los Muchachos Observatory, Canary Islands (SARA-ORM), at Kitt Peak National Observatory, Arizona (SARA-KP) and at Cerro-Tololo, Chile (SARA-CT). We obtain our observations using the Bessel {\it BVRI} filters mounted on SARA-ORM and SARA-KP, and the SDSS {\it griz} filters mounted on SARA-CT. The data analysis is executed following the standard aperture photometry technique employing IRAF v2.16.
The photometric standards utilized for calibration purposes were obtained from \citet{1992AJ....104..340L} for  and \citet{2002AJ....123.2121S}. The complete list of available standards is provided in Table 2\footnote{\url{http://www.eso.org/sci/observing/tools/standards/Landolt.html}} and Table 8\footnote{\url{https://classic.sdss.org/dr7/algorithms/standardstars/tab08.dat.txt}} of the respective papers. We selected standards from these tables which were approximately at similar airmass as our targets for each observation.
To convert the photometry to the AB system, we apply the standard Vega-AB corrections\footnote{\url{http://www.astronomy.ohio-state.edu/~martini/usefuldata.html}\label{note:note1}}. The foreground Galactic extinction correction is applied using the calculations from \citet{2011ApJ...737..103S}. The fluxes are extracted using the canonical flux to magnitude conversion\textsuperscript{\ref{note:note1}}, $\rm F_{\nu} = 10^{-(m_{AB}+48.6)/2.5}$. The final magnitudes in the AB system are reported in Table~\ref{tab:sara}.  

\section{Results}\label{sec:result}

\subsection{Soft X-ray long term variability}\label{sec:softvar}
Blazars are known to be variable sources in all energy bands \citep[e.g.][]{2010ApJ...722..520A,2013ApJS..207...28S,2016A&A...591A..21M,2017MNRAS.466.3309R,Jiang_2018,2018Ap&SS.363..167M,2019ApJ...871..211P, 2019arXiv190209077L}. Importantly, for powerful FSRQs in which, according to leptonic models, the EC dominates the high-energy part of the SED, most of the X-ray emission originates from the low-energy electron population \citep{2007MNRAS.382L..82G}. In this scenario, since low-energy electron cooling time is expected to be longer with respect to the high-energy electron one ($\rm t_{\rm cool}\propto \rm \gamma^{-1}$, where \gm~is the random Lorentz factor of the electrons, see \citealp{2013LNP...873.....G}), this radiation is not expected to vary significantly and on long timescales \citep[e.g.,][]{2009ApJ...697L..81B,2016MNRAS.462.1542S,2019ApJ...871..101P}. Detecting spectral variability in the soft X-ray (0.3-15\,keV) band may be indicative of a change of emission region location \citep[see][]{2015ApJ...807..167T}, hence magnetic field strength and radiative energy densities, as well as bulk Lorentz factor \citep{2007MNRAS.382L..82G}. It could also possibly be related to a change of the injection power \citep[see][]{2016MNRAS.462.1542S}, or be indicative of underlying acceleration processes such as shocks or magnetic reconnection \citep[e.g.,][]{2009ApJ...704...38S, 2008bves.confE..14M, 2016frap.confE..56D, 2019MNRAS.482...65C}.

Motivated to study such property for high-redshift blazars, we looked for more sensitive observations in the soft X-ray band.
Two of our sources, J151002$+$570243 and J212912$-$153841, have both \xmm and {\it Chandra} data\footnote{The observations of J151002+570243 were taken on 2002-05-11 by \xmm~and on 2001-06-10 by {\it Chandra} while the ones of J212912-153841 on 2001-05-01 by \xmm~and on 1999-11-16 by {\it Chandra}.}, in addition to XRT. In order to test the presence/absence of spectral variability, we used \texttt{XSPEC} to separately fit all three observations. To consistently compare the results, we used a simple absorbed power-law model, with absorption ($N_{\rm H}$) fixed at Galactic value, and we chose an energy range where all telescopes overlap, from 0.5 to 7\,keV. 
Performing such exercise leads us to the result that J151002$+$570243 does not show significant spectral or flux variability, and all parameters are consistent within the errors. On the other hand, J212912$-$153841 shows both flux and index variability ($>1\sigma$) in all three observations.  
In the case of J151002$+$570243, we are able to perform joint fit of {\it XMM-Newton} and {\it Chandra} observations with XRT ones (see Section~\ref{sec:xray_spec}), assuming that the flux and photon index detected represent the typical status of the source. 
The same is not applicable to J212912$-$153841, for which we do find variability. More detailed study in the soft X-ray band would be required to investigate possible scenarios explaining this property, such as change in the injected power or emission region location. 
For sources with multiple XRT observations, light-curves show no significant variability (see Section~\ref{sec:swift}). Therefore we are able to combine all XRT available data to increase the significance of the detection.

\begin{table*}[t!]
\scriptsize
 \begin{center}
 \caption{Table of X-ray Best-fit Spectral Parameters}\label{tab:results}
 \begin{tabular}{ +c -c -c -c -c -c -c -c -c -c } 
 \hline
 \multicolumn{10}{c}{J013126$-$100931} \\
 \hline
 Model & \multicolumn{5}{c}{Best-fit parameters} & Flux \tablenotemark{a}& $\chi^2$/d.o.f. & F-stat \tablenotemark{b}& $p$-values \tablenotemark{c}\\
  \multirow{2}{*}{PL}& $\Gamma$ & & & & & & & & \\ 
    & $1.40\pm 0.05$ & & & & & $(1.08\pm{0.05})\times10^{-11}$ & 192.61/179 & & \\
  \multirow{2}{*}{BPL}& $\Gamma_1$ & $\Gamma_2$ & $\rm E_b (\rm keV) $ & & & & & \\
     & $1.13_{-0.3}^{+0.1}$ & $1.52\pm0.1$ & $5.3_{-3.4}^{+2.3}$ & & & $9.50^{+0.5}_{-0.6}\times10^{-12}$ & 175.22/177 & 8.78 & $2\times10^{-4}$ \\
\rowstyle{\color{magenta}} 
 {\bf \multirow{2}{*}{LP} \tablenotemark{d}} & $\alpha$ & $\beta$ & $\rm E_0$ & & & & & &\\
\rowstyle{\color{magenta}} 
    & $1.02\pm 0.17$ & $0.23\pm0.1$ & 1(scale) & & & $(9.01\pm0.5)\times10^{-12}$ & 176.29/178 & 16.47 & $7\times10^{-5}$ \\
  \multirow{2}{*}{PL+ABS} & $\Gamma$ & $\rm N_H(intr.)$ & & & & & & \\
         & $1.45\pm 0.06$ & $4.52_{-2.7}^{+3.6}\times10^{22}$ & & & & $1.04^{+0.04}_{-0.30}\times10^{-11}$ & 183.46/178 & 8.77 & $3\times10^{-3}$ \\
 \multirow{2}{*}{BPL+ABS} & $\Gamma_1$ & $\Gamma_2$ & $\rm E_b (\rm keV) $ & $\rm N_H(intr.)$ & & & & \\
          & $1.18_{-0.18}^{+0.16}$ & $1.56\pm{0.10}$ & $6.23_{-3.35}^{+1.91}$ & $0.63_{-0.58}^{+3.39}\times10^{22}$ & & $9.45_{-0.72}^{+0.46}\times10^{-12}$ & 174.83/176 & 5.96 & $6\times10^{-4}$ \\
  \hline
 \multicolumn{10}{c}{J064632$+$445116} \\
 \hline
  Model & \multicolumn{5}{c}{Best-fit parameters} & Flux \tablenotemark{a}& $\chi^2$/d.o.f. & F-stat \tablenotemark{b}& $p$-values \tablenotemark{c}\\
\rowstyle{\color{magenta}} 
  {\bf \multirow{2}{*}{PL}} \tablenotemark{d}& $\Gamma$ & & & & & & & & \\ 
\rowstyle{\color{magenta}} 
    & $1.63\pm0.07$ & & & & & $2.27^{+0.12}_{-0.17}\times10^{-12}$ & 37.82/48 & & \\
  \multirow{2}{*}{BPL}& $\Gamma_1$ & $\Gamma_2$ & $\rm E_b (\rm keV)$ & & & & & \\
     & $1.51_{-0.14}^{+0.12}$ & $1.76^{+0.3}_{-0.09}$ & $4.11^{+14.6}_{-2.41}$ & & & $2.06^{+0.17}_{-0.45}\times10^{-12}$ & 34.54/46 & 2.18 & 0.12 \\
  \multirow{2}{*}{LP}& $\alpha$ & $\beta$ & $\rm E_0$ & & & & & & \\
    & $1.48^{+0.1}_{-0.2}$ & $0.11^{+0.17}_{-0.15}$ & 1(scale) & & & $2.04^{+0.22}_{-0.31}\times10^{-12}$ & 36.63/47 & 1.52 & 0.22 \\ 
 \hline
 \multicolumn{10}{c}{J151002$+$570243} \\
 \hline
  Model & \multicolumn{5}{c}{Best-fit parameters} & Flux \tablenotemark{a}& $\chi^2/$d.o.f. & F-stat \tablenotemark{b} & $p$-values \tablenotemark{c}\\
  \multirow{2}{*}{PL}& $\Gamma$ & & & & & & & &\\ 
    & $1.47\pm 0.05$ & & & & & $1.95^{+0.15}_{-0.09}\times10^{-12}$ & 391.61/276 & & \\
  \multirow{2}{*}{BPL}& $\Gamma_1$ & $\Gamma_2$ & $\rm E_b (\rm keV)$ & & & & & \\
    & $1.41\pm0.02$ & $0.98\pm{0.20}$ & $6.25^{+2.15}_{-1.10}$ & & & $(3.36\pm{0.50})\times10^{-12}$ & 381.27/274 & 3.71 & $2\times10^{-2}$ \\
  \multirow{2}{*}{LP}& $\alpha$ & $\beta$ & $\rm E_0$ & & & & & &\\
    & $1.44\pm 0.05$ & $-0.02\pm0.08$ & 1 (scale) & & & $(2.05\pm{0.30})\times10^{-12}$ & 388.32/275 & 2.32 & $10^{-1}$ \\
  \multirow{2}{*}{PL+ABS} \tablenotemark{d} & $\Gamma$ & & & $\rm N_H(intr.)$ & & & & & \\
   & $1.48\pm0.04$ & &  & $(0.73\pm0.31)\times10^{22}$ & & $(1.78\pm{0.21})\times10^{-12}$ & 376.07/275 & 5.38 & $8\times10^{-4}$ \\
\rowstyle{\color{magenta}} 
 {\bf \multirow{2}{*}{BPL+ABS}} \tablenotemark{d} & $\Gamma_1$ & $\Gamma_2$ & $\rm E_b (\rm keV)$ & $\rm N_H(intr.)$ & & & & & \\
\rowstyle{\color{magenta}} 
  & $1.52\pm0.05$ & $0.93_{-0.7}^{+1.17}$ & $6.02_{-1.7}^{+1.1}$ & $(0.86\pm0.32)\times10^{22}$ & & $(3.26\pm{0.5})\times10^{-12}$ & 360.27/273 & 7.91 & $4\times10^{-5}$ \\
\rowstyle{\color{magenta}} 
   {\bf \multirow{2}{*}{2BPL} \tablenotemark{d}} & $\Gamma_1$ & $\Gamma_2$ & $\Gamma_3$ & $\rm E_{b1} (\rm keV)$ & $\rm E_{b2} (\rm keV)$ & & & &\\
\rowstyle{\color{magenta}} 
    & $1.02_{-0.34}^{+0.27}$ & $1.49\pm0.04$ & $0.93\pm0.20$ & $0.85^{+0.28}_{-0.10}$ & $6.09_{-1.50}^{+1.24}$ & $(3.41\pm{0.5})\times10^{-12}$ & 355.18/272 & 6.97 & $8\times10^{-5}$ \\
 \hline
 \multicolumn{10}{c}{J212912$-$153841} \\
 \hline
  Model & \multicolumn{5}{c}{Best-fit parameters} & Flux \tablenotemark{a}& $\chi^2/$d.o.f. & F-stat \tablenotemark{b}& $p$-values \tablenotemark{c}\\
  \multirow{2}{*}{PL}& $\Gamma$ & & & & & & & \\ 
    & $1.38\pm 0.02$ & & & & & $3.53^{+0.08}_{-0.07}\times10^{-11}$ & 813.92/631 & & \\
  \multirow{2}{*}{BPL}& $\Gamma_1$ & $\Gamma_2$ & $\rm E_b (\rm keV)$ & & & & & & \\
     & $0.91\pm 0.09$ & $1.56\pm0.03$ & $1.75^{+0.30}_{-0.17}$ & & & $3.00^{+0.06}_{-0.08}\times10^{-11}$ &  542.37/629 & 157.462 & $10^{-56}$ \\
  \multirow{2}{*}{LP}& $\alpha$ & $\beta$ & $\rm E_0$ & & & & \\
    & $1.10\pm0.04$ & $0.25\pm0.20$ & 1(scale) & & & $2.57^{+0.08}_{-0.07}\times10^{-11}$ & 569.69/630 & 270.085 & $10^{-50}$ \\
  \multirow{2}{*}{PL+ABS} & $\Gamma$ & $\rm N_H(intr.)$ & & & & & & \\
         & $1.51\pm0.02$ & $2.16\pm0.33\times10^{22}$ & & & & $(3.09\pm{0.09})\times10^{-11}$ & 560.98/630 & 284.06 & $10^{-53}$\\
\rowstyle{\color{magenta}} 
{\bf \multirow{2}{*}{BPL+ABS} \tablenotemark{d}} & $\Gamma_1$ & $\Gamma_2$ & $\rm E_b (\rm keV)$ & $\rm N_H(intr.)$ & & & & &\\
\rowstyle{\color{magenta}} 
   & $1.22_{-0.13}^{+0.09}$ & $1.57\pm0.03$ & $2.21\pm0.35$ & $0.98_{-0.46}^{+24}\times10^{22}$ & & $2.95^{+0.04}_{-0.10}\times10^{-11}$ & 528.13/628 & 113.278 &  $10^{-58}$ \\
\hline  
 \multicolumn{10}{c}{J020346$+$113445} \\
 \hline
  Model & \multicolumn{5}{c}{Best-fit parameters} & Flux \tablenotemark{a}& C-stat/d.o.f. & &\\
  \multirow{2}{*}{PL} & $\Gamma$ & & & & & & & & \\ 
     & $1.59\pm0.13$ & & & & & $3.26^{+0.25}_{-0.33}\times10^{-12}$ & 376.90/456 & & \\
 \hline  
 \multicolumn{10}{c}{J135406$-$020603} \\
 \hline
  Model & \multicolumn{5}{c}{Best fit parameters} & Flux \tablenotemark{a}& C-stat/d.o.f. & & \\
  \multirow{2}{*}{PL} & $\Gamma$ & & & & & & & & \\ 
     & $1.32\pm0.16$ & & & & & $1.99^{+0.19}_{-0.32}\times10^{-12}$ & 398.68/477 & & \\
  \hline
\end{tabular}
  \begin{minipage}{\textwidth}
 \tablenotetext{} {\bf Notes:}
 \tablenotetext{a}{Observed flux in units of $10^{-12}$\ergflux~in the $0.3-79$\,keV energy band. The errors are at the 90\% level of confidence for one parameter of interest and the fluxes are corrected for Galactic absorption.}
 \tablenotetext{b,c}{$\quad$The F-statistic and $p$-values are calculated for the different models with respect to the power-law case.}
 \tablenotetext{d}{Best-fit model extracted from the analysis is highlighted in magenta.}
 \tablenotetext{}{}
\end{minipage}%
\end{center}
\end{table*}
\begin{figure}
\centering
\includegraphics[width=0.45\textwidth]{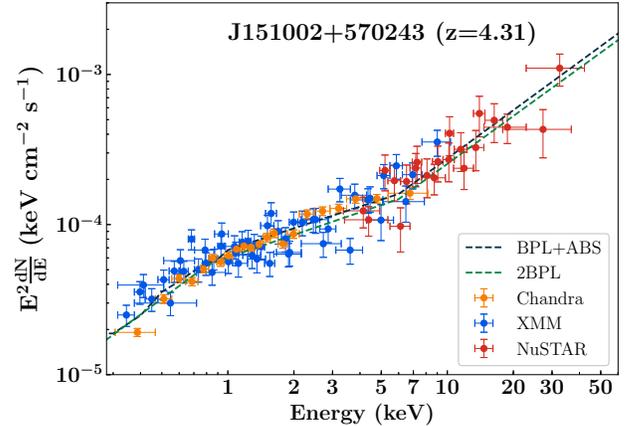}
\caption{Combined {\it Chandra}, {\it XMM}, {\it Swift}-XRT and \nustar~spectrum of J151002$+$570243. The best-fit shapes obtained using XSPEC are a double broken power-law (2BPL) as well as a broken power-law with excess absorption (BPL+ABS). The low-energy break can be attributed to either an intrinsic property of the underlying electron distribution, or to an intervening absorber in the IGM. The softening (between $\sim0.8$\,keV and $\sim6$\,keV) and re-hardening ($>7$\,keV) of the spectrum can be explained within the one-zone leptonic model (see Section~\ref{sec:model}).\label{fig:TXS_xray}}
\end{figure}
\begin{table}[t!]
\centering 
\caption{Table of Black-Hole Masses, derived both from spectroscopic approach and SED modeling.}\label{tab:mass}
 \begin{tabular}{ c | c c } 
 \hline
                         &  $M_{\rm BH, SED}(M_{\sun})$ & $M_{\rm BH, spectroscopy}$ ($M_{\sun}$) \\
\hline
        J013126$-$100931 & $6.0\times10^{9}$ & ... \\
        J020346$+$113445 & $3.0\times10^{9}$ & ... \\ 
        J064632$+$445116 & $1.5\times10^{9}$ & $1.3 \times 10^{9}$\\
        J135406$-$020603 & $1.0\times10^{9}$ & $8.9 \times 10^{8}$\\
        J151002$+$570243 & $6.5\times 10^{9}$& $3.1 \times 10^{8}$\\
        J212912$-$153841 & $5.0\times10^{9}$ & $6.3\times10^{9}$ \\
 \hline
\end{tabular}
\end{table}

\subsection{X-ray Spectral Analysis}\label{sec:xray_spec}
We fit the soft X-ray (0.3-15\,keV) and \nustar~FPMA and FPMB (3-79\,keV) spectra using XSPEC. We include Galactic absorption ($N_{\rm H}$), using the Galactic neutral hydrogen column densities from \citet[][]{2005A&A...440..775K}.
We use the C-statistics \citep[][]{1979ApJ...228..939C} for sources with low soft X-ray counts ($<30\rm\,cts$), otherwise we employ standard $\chi^2$ statistics.  
Sources with good X-ray signal are: J013126$-$100931, J064632$+$445116, J151002$+$570243 and J212912$-$153841. 
High-redshift blazars have been found to show spectral softening at soft X-rays \citep{2005MNRAS.364..195P,2011ApJ...738...53S,2013ApJ...774...29E}, which has been proposed to be linked to either a property of the intrinsic electron population underlying the emission or to the presence of an intervening absorber residing in the interstellar Galactic medium \citep[IGM, see][]{2018A&A...616A.170A}, or both. In order to test this scenario, at first these objects are fitted with a simple power-law model along with two curved ones, broken power law and log-parabola, all with absorption fixed at the Galactic value. 
Then, if a curved model is favored over a simple power-law one (highlighting the presence of a softening of the soft X-ray part of the spectrum), we also test the possibility of an intrinsic absorber. Therefore, we further fit these objects with both a power-law and a broken power-law model including a cold absorber (\texttt{ztbabs} in \texttt{XSPEC}).
All X-ray spectral fit parameters are provided in Table~\ref{tab:results} and the fits and residuals for each fitted model can be found in Appendix~\ref{app:A}. The main results are listed below:
\begin{enumerate} 
\item Fits' residuals and F-test results showed that both J013126$-$100931 and J212912$-$153841 favored a curved model. Fits including the absorber give a reduced $\chi^2$ close to one and of the same order as the one found for the log-parabola/broken power-law case (see Table~\ref{tab:results}). Results using the F-test favor a log-parabola fit for J013126$-$100931, while J212912$-$153841 is best-fitted by a broken power-law model with intrinsic absorber. The radiating electron population is therefore likely in both cases to possess an intrinsic break at low energies, with J212912$-$153841 also showing the signature of an intervening absorbing medium.   
\item The source J151002$+$570243 represents an exception. Analysing residuals and F-test results, the broken power-law model plus absorber is favored over the broken power-law ($p$-value$=8\times10^{-5}$) or power-law with absorber ($p$-value$=2\times10^{-3}$) ones. 
As can be seen from Figure~\ref{fig:TXS_xray}, the spectrum indeed shows a break around 0.8\,keV, after which it becomes softer, and a second break around 6\,keV, after which it hardens again.
While the second break can be explained via detailed modeling of the jet properties (see Section~\ref{sec:disc}), the first one can hint to either the presence of an intrinsic absorber or a break in the electron population itself. To check whether the second option is viable, we further test a double broken power-law model\footnote{A similar test has been performed for J212912$-$153841, but a double broken power-law does not result in a reliable fit, hence it is excluded from Table~\ref{tab:results}.} (\texttt{bkn2po} in \texttt{XSPEC}). 
The fit is again preferred over the broken power-law ($p$-value$=6\times10^{-5}$) or power-law with absorber ($p$-value$=10^{-3}$) models. Both the broken power-law plus intrinsic absorber and double broken power-law fits give similar reduced $\chi^2$ values, and the F-test does not favor one over the other.  
\item For J020346+113445 and J135406−020603, due to low soft X-ray statistics, a simple power-law model with absorption fixed at the Galactic value is employed. 
\end{enumerate} 
In order to cross calibrate the instruments, we include a multiplicative constant factor, fixing it equal to 1 for FPMA but leaving it free to vary for FPMB and soft X-ray instruments.
For all sources with good X-ray statistics, the constants between instruments are of the order of $\sim$1-10\% from the fixed FPMA value. Instead, for the ones with poor soft X-ray statistics the difference for FPMB is in the range 0.6-10 \%, while a larger offset in the range 10-30\% is found for \swift-XRT.

\subsection{Central black hole mass}\label{sec:massbh}
Black hole masses are computed following two approaches. The first is through single epoch optical spectroscopy \citep[e.g.,][]{2011ApJS..194...45S}, assuming a virialized BLR and adopting emission lines parameters such as full width at half maximum (FWHM), luminosity and continuum emission. The errors associated with this method are of the order of $\sim 0.4$ dex \citep[e.g.,][]{2006ApJ...641..689V,2011ApJS..194...45S}. 
The second is by modeling the accretion disk using a standard \citet[][]{1973A&A....24..337S} model. Blazars' accretion disk emission falls in the IR-UV band and, if measurable, it enables us to derive the black hole mass, even when optical spectra are not available. High-redshift blazars, thanks to the shift of the low-energy SED peak, unveil this emission and are therefore ideal sources for applying this method. With a good IR-UV coverage, the uncertainty associated with this approach are of about a factor of 2 and the results have been shown to be in good agreement with the virial method \citep[see e.g.][]{2013A&A...560A..28C,2015MNRAS.448.1060G,2017ApJ...851...33P}.
The accretion disk spectral energy distribution is assumed to be a multi-color blackbody. The flux density profile of this emission is given by the following \citep[][]{2002apa..book.....F}:

 \begin{equation}
 \label{eq:SS_discflux}
 F_\nu = \nu^3\frac{4\pi h \cos \theta_{\rm v} }{c^2 {D}^2}\int_{R_{\rm in}}^{R_{\rm out}}\frac{R\,{\rm d}R}{e^{h\nu/kT(R)}-1},
\end{equation}
where $k$, $c$ and $h$ are the Boltzmann constant, the speed of light and the Plank constant, respectively, $D$ is the luminosity distance, $\theta_{\rm v}$ is the jet viewing angle and $R_{\rm in}$ and $R_{\rm out}$ are the inner and outer disk radii. We take $R_{\rm in} = 3R_{\rm Sch}$ and $R_{\rm out}=500 R_{\rm Sch}$ ($R_{\rm Sch}$ is the Schwarzschild radius).
The radial temperature profile is given by:

\begin{equation}
T(R)\, =\, {  3 R_{\rm Sch}  L_{\rm disk }  \over 16 \pi\eta_{\rm a}\sigma_{\rm SB} R^3 }  
\left[ 1- \left( {3 R_{\rm Sch} \over  R}\right)^{1/2} \right]^{1/4},
\end{equation}
where $\sigma_{\rm SB}$ is the Stefan-Boltzmann constant and $\eta_{\rm a}$ is the accretion efficiency (we consider $\eta_{\rm a}=0.1$).
\begin{figure*}
      \gridline{\fig{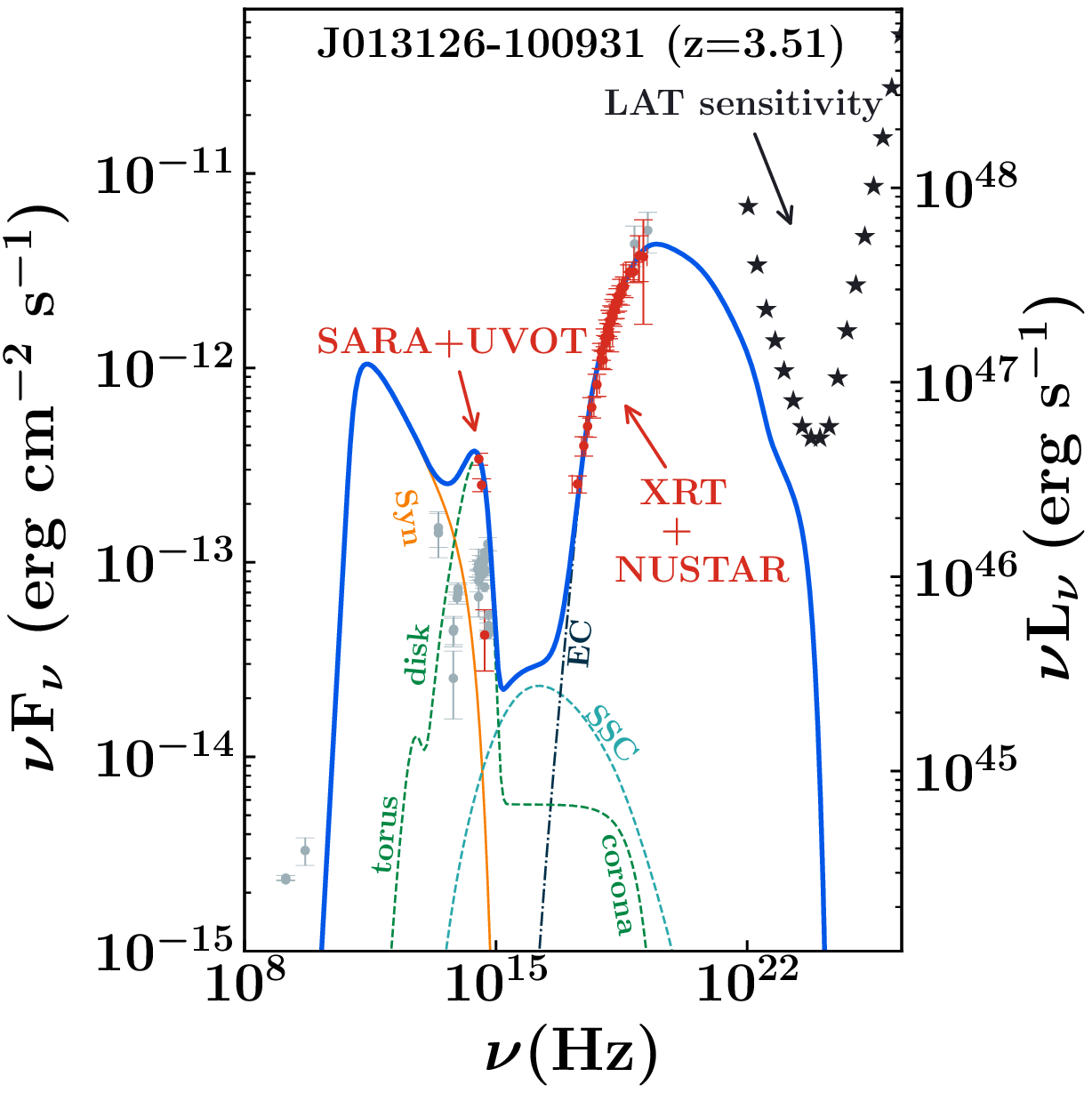}{0.33\textwidth}\
	  \fig{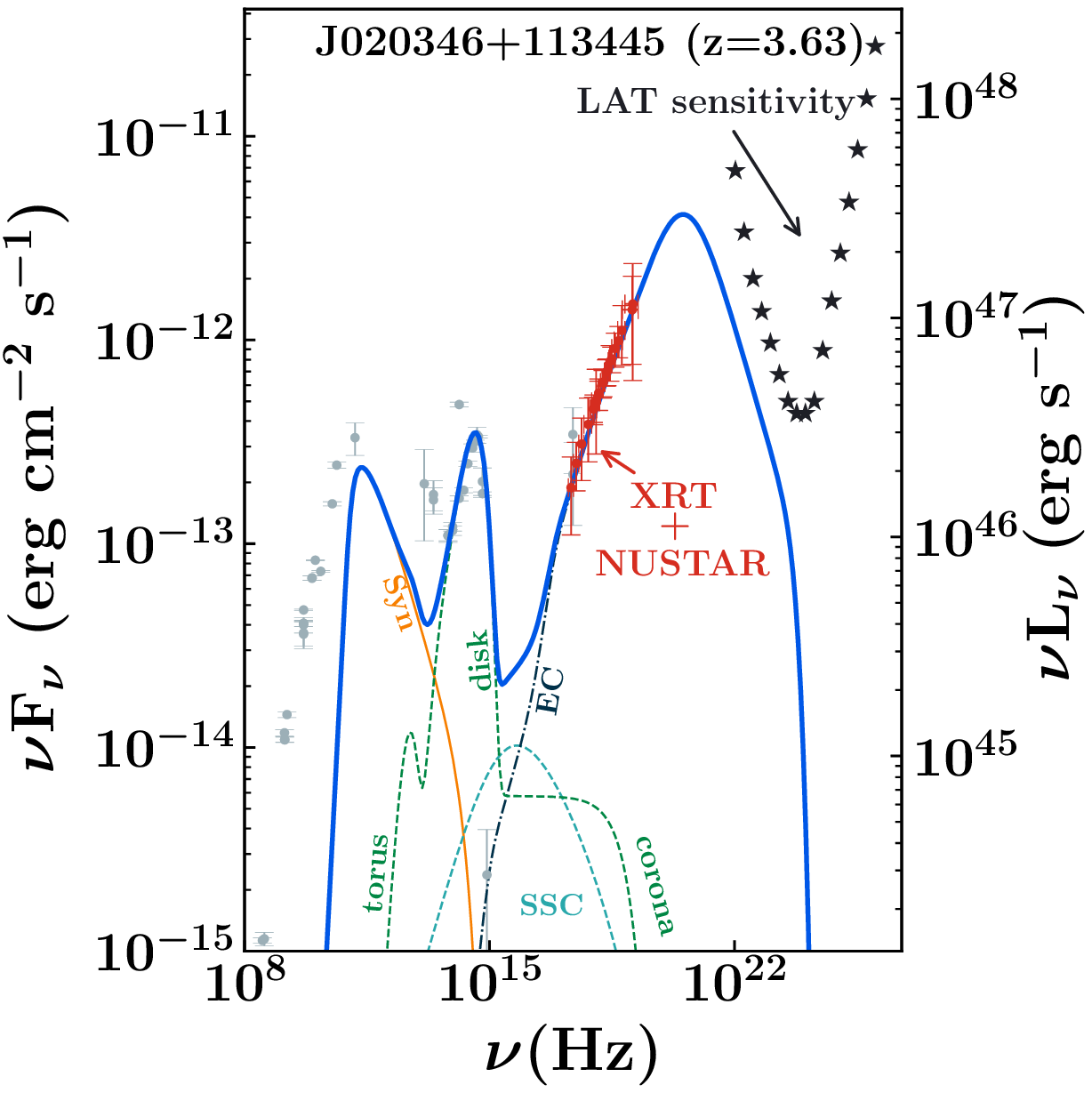}{0.33\textwidth}\
          \fig{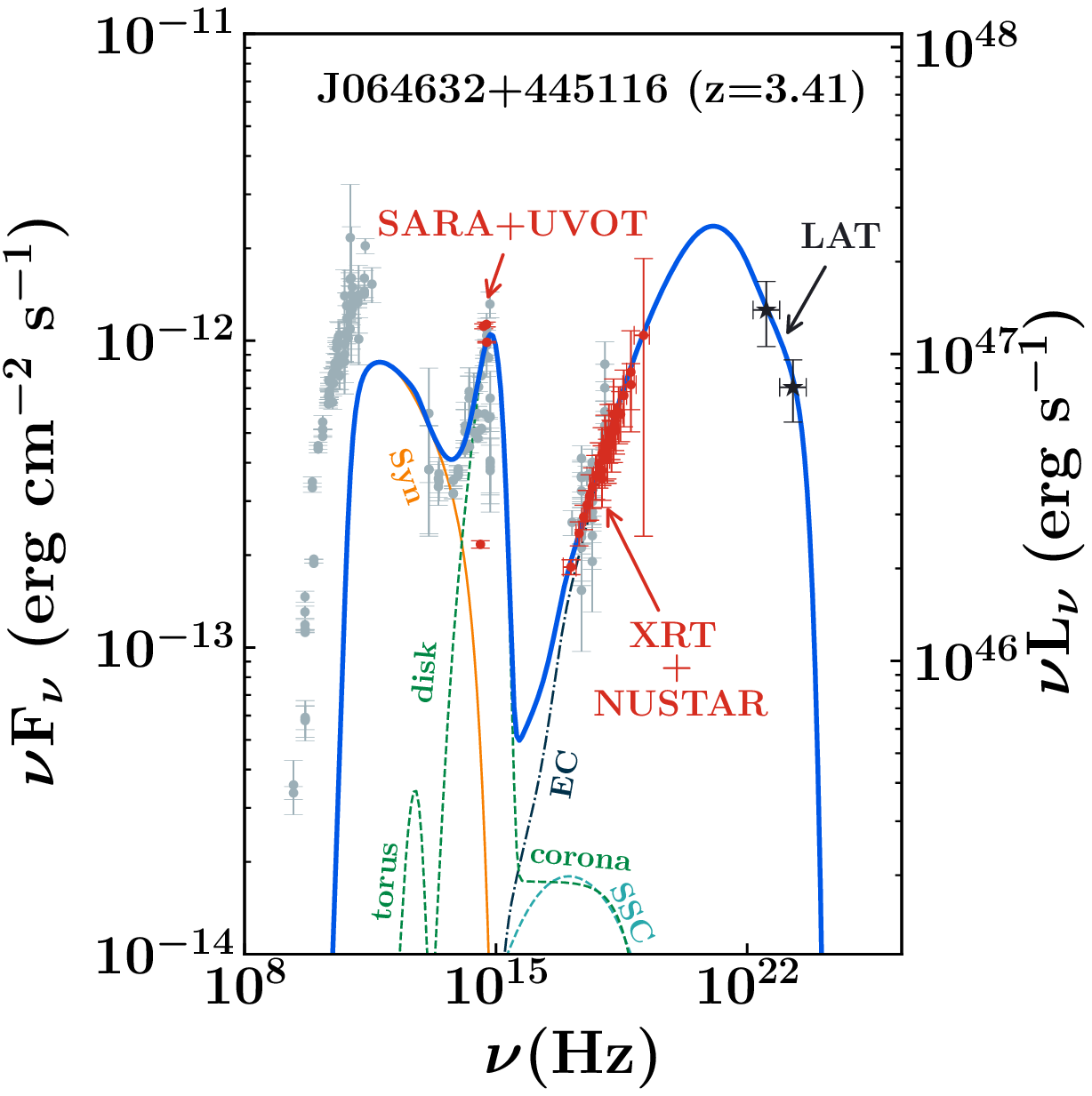}{0.33\textwidth}\
           }           

	\gridline{\fig{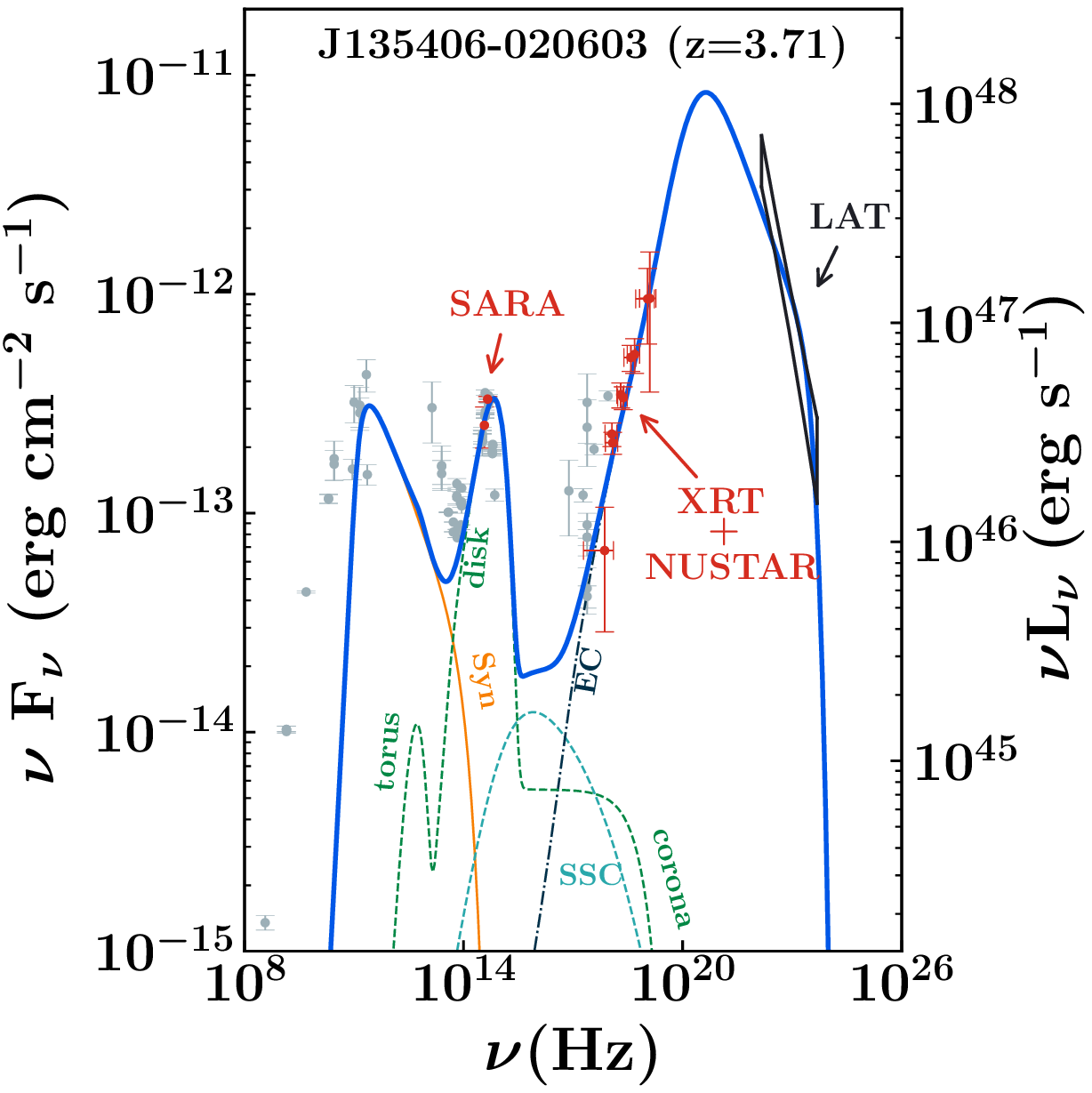}{0.33\textwidth}\
	  \fig{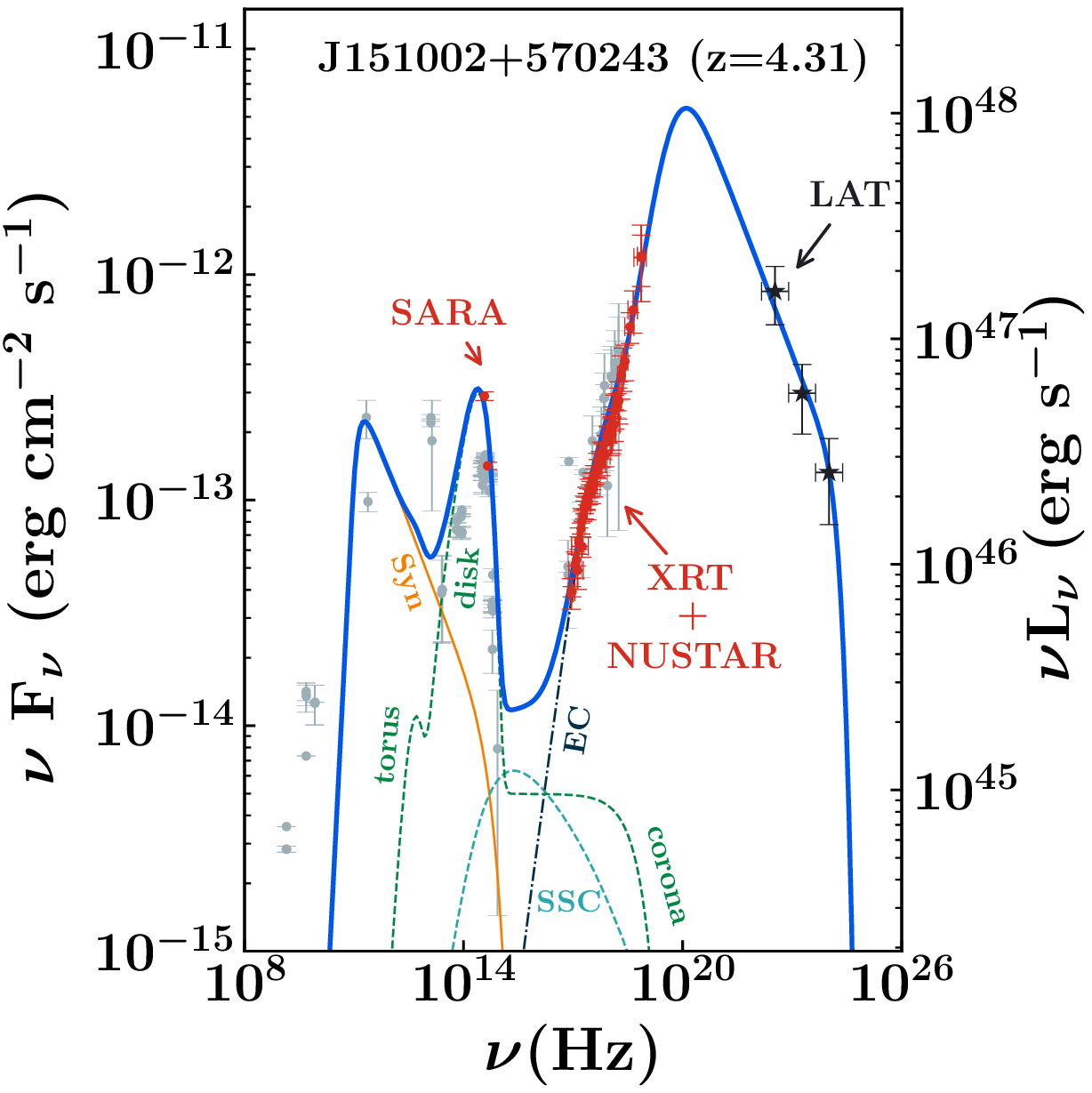}{0.33\textwidth}\
          \fig{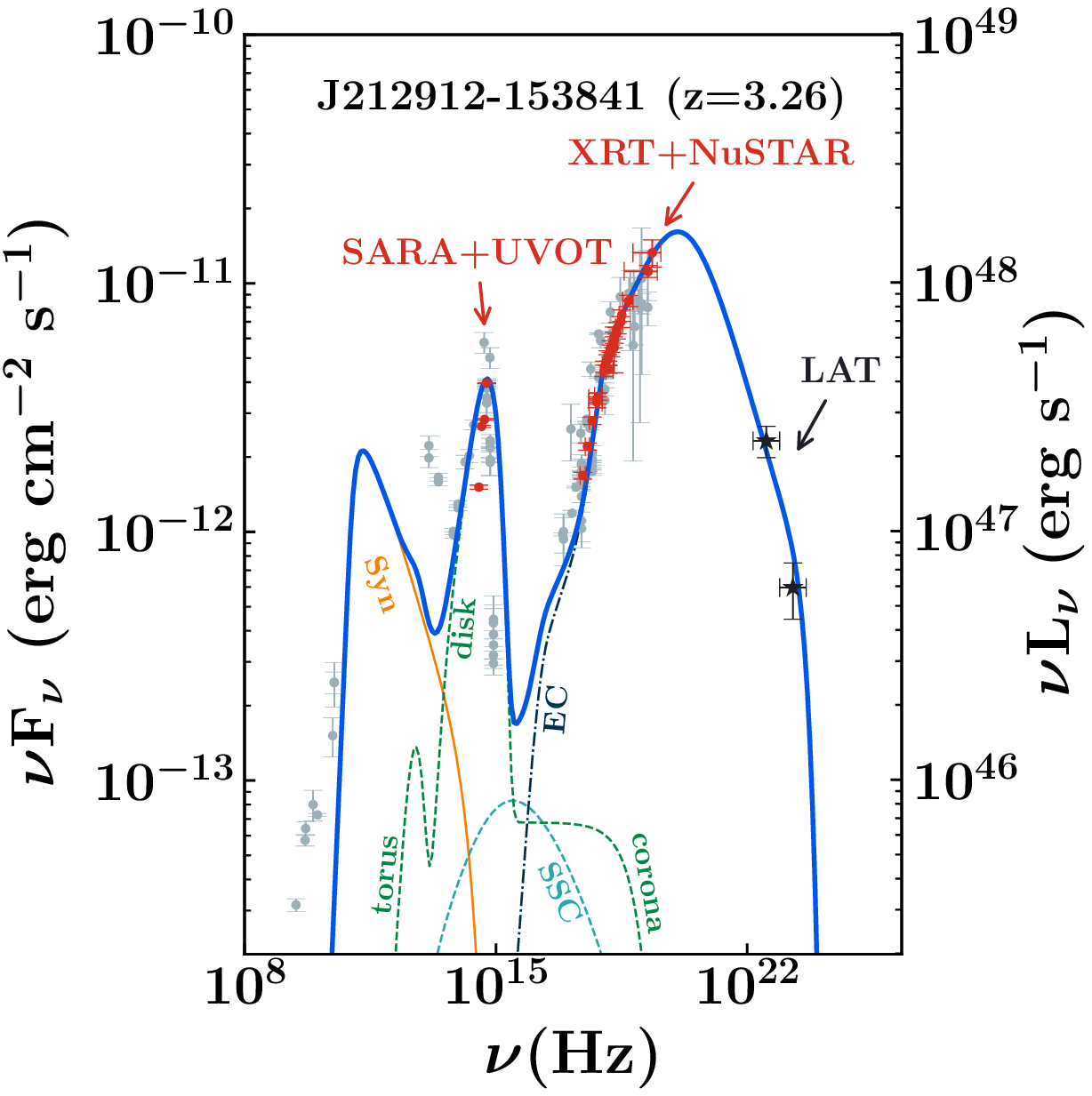}{0.33\textwidth}\
           }
	\caption{The broadband SED of six quasars using quasi-simultaneous SARA, \swift, \nustar, and \fermi-LAT data, modeled via the one zone leptonic emission model described in the text. Grey circles represent the archival data, the black stars are the \fermi~data extracted from \citealp{2017ApJ...837L...5A}, while the red points show our quasi-simultaneous observations. Since the LAT SED data points for J135406-020603 are not published in \citealp{2017ApJ...837L...5A}, we use the parameters listed in their paper to extract the bowtie plot showed here.  The lines are the modeling lines illustrating different components considered: the orange line is the synchrotron emission; the green dotted line includes the thermal emission from the torus, the disk and the corona; the light-blue dotted line represents the SSC component; the dark blue dotted line is the EC component. The blue solid line is the sum of all the other contributors and delineate our best fit model (see Section~\ref{sec:disc} for detailed discussion). For the optical data, when available, we plot the photometric data below the redshifted Ly-$\alpha$ frequency since above this limit the emission is significantly affected by absorption from intergalactic hydrogen \citep[e.g.][]{Songaila_2010,2018MNRAS.481.1320C}. \label{fig:sed}}
\end{figure*}
With the above formulation, there are only two free parameters:
the black hole mass ($M_{\rm BH}$) and the accretion rate ($\dot{M}_{\rm a}$). The latter is related to the intrinsic accretion luminosity via $L_{\rm disk}=\eta_{\rm a}\dot{M}_{\rm a}c^2$. If the peak of the disk emission is visible, it is possible to obtain $L_{\rm disk}$ from the observations. The only free parameter left, therefore, is $M_{\rm BH}$, which scales as $M_{\rm BH}\propto\nu_{\rm peak,disk}^{-2}{L_{\rm disk}}^{1/2}$ (where $\nu_{\rm peak,disk}$ is the peak of the disk emission, see \citealp{2013MNRAS.431..210C})
\begin{table*}[t!]
\caption{Summary of the parameters used/derived from the SED modeling of six MeV blazars shown in Figure~\ref{fig:sed}. A viewing angle of 3$^{\circ}$ is adopted for all of them.}\label{tab:sed_par}
\hspace{-1.5cm}
\resizebox{1.1\textwidth}{!}{
\begin{tabular}{lccccccc}
\tableline
\tableline
Parameter & J013126$-$100931  &  J020346$+$113445 & J064632$+$445116 & J135406$-$020603  & J151002$+$570243 & J212912$-$153841 \\ 
\tableline
\tableline
Slope of the particle distribution below the break energy ($p$)   & 1.5  & 1.8 & 2.05 & 0.9 & 0.5 & 2.1 \\      
Slope of the particle distribution above the break energy ($q$)   & 3.8  & 4.2 & 3.6 & 3.9  & 3.8  & 4.0 \\      
Magnetic field in Gauss ($B$)                                     & 0.5  & 0.9 & 1.0 & 1.2  & 1.2  & 0.8 \\      
Particle energy density in erg cm$^{-3}$ ($U_{e}$)                & 0.003 & 0.02& 0.004 & 0.01 & 0.01 & 0.009\\       
Bulk Lorentz factor ($\Gamma$)                                    & 15.8   & 10.0  & 14.0    & 10.0   & 11.0 & 10.0 \\      
Minimum Lorentz factor ($\gamma_{\rm min}$)                       & 12    & 1   & 1     & 1    & 1.7 & 4.0 \\      
Break Lorentz factor ($\gamma_{\rm break}$)                       & 74.07 & 83.85 & 131.22 & 59.80 & 32.42 & 60.32\\      
Maximum Lorentz factor ($\gamma_{\rm max}$)                       & 3e3   & 3e3   & 3e3   & 3e3 & 5e3 & 3e3 \\      
Dissipation distance in parsec ($R_{\rm diss}$, $R_{\rm Sch}$)& 0.43 (750) & 0.28 (1000)   & 0.57 (4000) & 0.25 (2700) & 0.31 (500)  & 0.98 (2050) \\
Size of the BLR in parsec ($R_{\rm BLR}$, $R_{\rm Sch}$) & 0.22 (399) & 0.23 (837) & 0.38 (2671) & 0.23  (2511) & 0.27 (438)  & 0.72 (1514)\\    
Accretion disk luminosity in log scale ($L_{\rm disk}$, erg s$^{-1}$)   & 46.70  & 46.74 & 47.15 & 46.74 & 46.85 & 47.71 \\       
Accretion disk luminosity in Eddington units ($L_{\rm disk}/L_{\rm Edd}$)   & 0.06 & 0.14 & 0.74 & 0.43 & 0.09 & 0.79 \\      
\tableline
Jet power in electrons in log scale ($P_{\rm e}$, erg s$^{-1}$)    & 45.19 & 45.29 & 45.37  & 44.93 & 45.13 & 45.89\\      
Jet power in magnetic field in log scale ($P_{\rm B}$), erg s$^{-1}$   & 45.61 & 45.37 & 46.36 & 45.53 & 45.77 & 46.33\\      
Radiative jet power in log scale ($P_{\rm r}$, erg s$^{-1}$)       & 45.49 & 46.11 & 46.27 & 46.59 & 46.37 & 46.70 \\      
Jet power in protons in log scale ($P_{\rm p}$, erg s$^{-1}$)      & 46.90 & 47.78 & 47.98 & 46.89 & 47.12 & 48.11 \\      
Total jet power in log scale ($P_{\rm TOT}$, erg s$^{-1}$)         & 46.93 & 47.78 & 47.99 & 46.91  & 47.14 & 48.12  \\      
\tableline
\tableline
\end{tabular}
}
\end{table*}

For our sources, we looked into the literature to find the black hole masses derived through the first method, i.e.\ optical spectroscopy, then we performed disk modeling to extract them through the second. 
Both masses are reported in Table~\ref{tab:mass}. They are found to be in reasonable accordance, and match within a factor of 2, apart for J151002$+$570243, whose difference cannot be accounted for by considering the associated errors. 
In fact, we note that the SARA magnitudes detected for this source are larger ($\sim$ a factor of 2 higher in flux) than the average archival ones. This could be linked to the fact that we are observing the source in a high state \citep[see e.g.,][]{2009ApJ...698..895K,2014ApJ...783..105R,2018ApJ...854..160R}, hence our measurement of the disk luminosity could be higher than the low/average one. However, the peak frequency of the emission is consistent with what found in the archival data, therefore the estimate on the black hole mass is not strongly affected. In the case of J013126-100931 we find that both SARA magnitudes are larger\footnote{The source J013126-100931 is present in the {\it Catalina Sky Survey} (CSS, \citealp{2009ApJ...696..870D}), showing a range of magnitudes compatible with that found by our observations.} and the peak of the emission is at lower frequencies with respect to archival data. This could lead to an overestimate on the black hole mass, although the value would still be within the errors associated with the modeling approach. 
Longer term optical monitoring would be needed to study variability of the disk emission.

\subsection{Broadband SED modeling}\label{sec:model}
We reproduce the broadband SED of all six sources following a simple one-zone leptonic emission model \citep[see,][]{2009MNRAS.397..985G, 2009ApJ...692...32D} and here we explain it briefly.
We assume that radiating electrons are confined in a spherical region
canvassing the full jet cross-section. This region is considered to be located at a distance $R_{\rm diss}$ from the central black hole, moving with bulk Lorentz factor $\Gamma$. The jet semi-opening angle is taken to be 0.1 rad.  
The energy distribution of relativistic electrons is modeled as a broken power law of the following type:
 \begin{equation}
 N(\gamma)  \, \propto \, { (\gamma_{\rm break})^{-p} \over
(\gamma/\gamma_{\rm break})^{p} + (\gamma/\gamma_{\rm break})^{q}}. \label{eq:shape}
\end{equation}
where {\it p} and {\it q} are, respectively, the slopes of the distribution before and after the break energy, $\gamma_{\rm break}$.
Relativistic electrons, immersed in a randomly oriented but uniform magnetic field ($B$), emit synchrotron radiation and, in presence of a photon field, they also undergo IC process. We consider the 
sources of radiation for IC to be either the same photons emitted via synchrotron or radiation fields external to the jet. 
For the latter, following \citet{2009MNRAS.397..985G}, we take into account (1) the accretion disk emission, (2) the X-ray corona located above the accretion disk, 
(3) the Broad Line Region (BLR), (4) the dusty torus. The corona is assumed to have a cut-off power-law spectral distribution, and to reprocess 30\% of the disk luminosity. The BLR is considered to be a spherical shell located at a distance $R_{\rm BLR} = 10^{17} L^{1/2}_{\rm disk,45}$ cm ($L_{\rm disk,45}$ being the accretion disk luminosity in units of 10$^{45}$ \lum) from the central engine, and reemitting 10\% of $L_{\rm disk}$ \citep{2006ApJ...644..133B,2007ApJ...659..997K,2008MNRAS.386..945T}. The torus is also assumed to be a spherical shell, located further away at $R_{\rm TORUS} = 10^{18} L^{1/2}_{\rm disk,45}$ cm and reprocessing 50\% of $L_{\rm disk}$ \citep{1991MNRAS.252..586L,2002ApJ...570L...9N,2007ApJ...660..117C}. The spectral shapes of both the BLR and the torus are considered to be blackbodies 
peaking, respectively, at the Lyman-$\alpha$ frequency and at the characteristic torus temperature ($T_{\rm TORUS}=300\,\rm K$). Calculating the 
radiative energy densities of all four components, we are able to extract the IC spectra (both SSC and EC).
In order to calculate the total jet power, we make use of the following assumptions: the kinetic power of the jet is carried only by protons, considered to be cold and therefore contributing only to the inertia of the jet; the number densities of relativistic electrons and protons are equal \citep[e.g.,][]{2008MNRAS.385..283C}.
The total jet power is evaluated as the sum of electrons ($P_{\rm e}$), protons ($P_{\rm p}$), magnetic field ($P_{\rm m}$) and radiation ($P_{\rm r}$) powers, which are all calculated. 
\begin{table*}[t!]
\begin{center}
\begin{minipage}{\textwidth}

\begin{minipage}{0.5\textwidth}
\resizebox{\textwidth}{!}{
\begin{tabular}{lcc}
\tableline
\tableline
\multicolumn{3}{c}{J064632$+$445116} \\
\tableline
\tableline
Parameter & \citet{2017ApJ...837L...5A} & This work \\ 
\tableline                                                         
\tableline
$p$   & 1.8 &  2.05 \\      
$q$   & 4.4 & 3.6 \\      
$B$ [Gauss]                                     & 2.1 & 1.0\\      
$U_{e}$erg cm$^{-3}$                & 0.009 & 0.004 \\       
$\Gamma$                                    & 12 & 14    \\      
$\gamma_{\rm min}$                       & 1 & 1     \\      
$\gamma_{\rm break}$                       & 72 & 131.22 \\      
$\gamma_{\rm max}$                       & $2\times10^3$ & $3\times10^3$ \\      
$R_{\rm diss} [\rm parsec]$                   & 0.25 & 0.57  \\     
$R_{\rm BLR} [\rm parsec]$                       & 0.37 & 0.38 \\    
\tableline
$P_{\rm e} [\rm erg~s^{-1}$, in log scale$]$      & 44.88 & 45.37  \\      
$P_{\rm B} [\rm erg~s^{-1}$, in log scale$]$ & 46.15  & 46.36 \\      
$P_{\rm r} [\rm erg~s^{-1}$, in log scale$]$         & 45.89 & 46.27 \\      
$P_{\rm p} [\rm erg~s^{-1}$, in log scale$]$        & 47.38 & 47.98 \\      
$P_{\rm TOT} [\rm erg~s^{-1}$, in log scale$]$           & 47.41 & 47.99\\
\tableline
\tableline
\end{tabular}
}
\captionof{table}{Table of comparison for the parameters used/derived from the SED modeling of J064632$+$445116 from \citet{2017ApJ...837L...5A} and this work.\label{tab:sed_par_nus}}
\end{minipage}
\hspace{0.2cm}
\begin{minipage}{0.48\textwidth}
\centering
\includegraphics[width=0.8\textwidth]{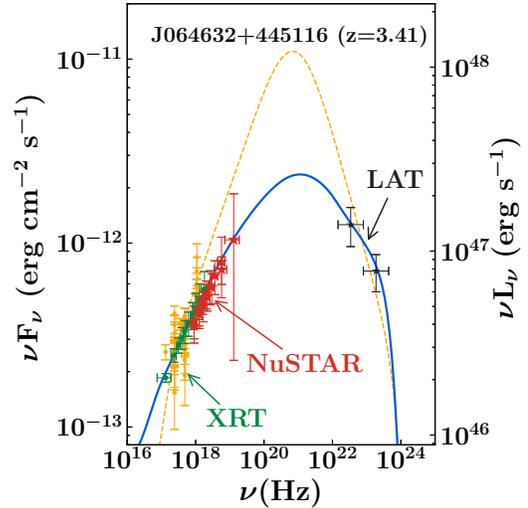}
\captionof{figure}{Zoomed high-energy SED of J064632+445116. The yellow/orange data points are the archival observations and the yellow/orange line is the SED model derived by \citealp{2017ApJ...837L...5A}. The blue line is the best fit model we obtained using combined soft X-ray (green), hard X-ray (red) and \gm-ray (black) data. We emphasize that the availability of \nustar~observations allow us to tightly constrain the rising part of the EC peak, enabling us to accurately infer the power of the jet and the underlying particle distribution shape (see Table~\ref{tab:sed_par_nus}).\label{fig:sed_par_nus}}
\end{minipage}
\end{minipage}
\end{center}
\end{table*}

The simultaneous multiwavelength coverage obtained for all our sources is crucial to accurately determine the input parameters to our model (i.e.\ $L_{\rm disk}$, $R_{\rm diss}$, $M_{\rm BH}$, $p$, $q$, $\Gamma$, $B$, low- and high-energy cut-off of the electron distribution ($\gamma_{\rm min}$, $\gamma_{\rm max}$)). Here we list some guidelines behind our choice:
\begin{itemize}
\item Thermal component: optical data of these sources sample the accretion disk emission. It is therefore possible to gauge the peak of this radiation and hence the overall disk luminosity ($L_{\rm disk}\sim 2\nu_{\rm peak,disk}L(\nu_{\rm peak,disk})$). Following the discussion in Section~\ref{sec:massbh}, optical data also allow us to constrain $M_{\rm BH}$.  
\item Non-thermal component: blazars' emission region in this one-zone model is required to be compact and located within the BLR or the torus. As a consequence, synchrotron emission from this region is depleted by synchrotron self-absorption below $\sim 10^{12}\,\rm Hz$. It follows that the low-energy synchrotron component of the SED is likely to be produced by another region further away along the jet, therefore cannot be used to constrain the shape of the electron energy distribution. However, good quality X- and \gm-rays allow us to exactly determine peak of the IC emission and the shape of the non-thermal continuum, which is directly linked to the shape of the electron distribution. Hence, availability of both \nustar~and \fermi-LAT data are key to fix both $p$ and $q$ (see Equation~\ref{eq:shape}). Furthermore the interplay of $\Gamma$, $B$ and $R_{\rm diss}$ reflects on the level of the high-energy emission. Therefore, the X-to-\gm-ray available data allow us to accurately estimate these parameters. 
\end{itemize}
We report the parameters derived from the SED modeling in Table~\ref{tab:sed_par}. Figure \ref{fig:sed} shows the results for all six sources, which are discussed in the next Section. 
\begin{table*}[t!]
\begin{center}
\begin{minipage}{\textwidth}

\begin{minipage}{0.5\textwidth}
\resizebox{\textwidth}{!}{
\begin{tabular}{lcc}
\tableline
\tableline
\multicolumn{3}{c}{J151002+570243} \\
\tableline
\tableline
Parameter & \citet{2017ApJ...837L...5A} & This work \\ 
\tableline                                                         
\tableline
$p$  						& 1.8 & 0.5 \\      
$q$  						& 4.1 & 3.8 \\      
$B$ [Gauss]                                     & 1.4 & 1.2\\      
$U_{e}$erg cm$^{-3}$             		& 0.029 & 0.01 \\       
$\Gamma$                                   	& 11 & 11    \\      
$\gamma_{\rm min}$                      	& 1 & 1.7     \\      
$\gamma_{\rm break}$                      	& 82 & 32.42 \\      
$\gamma_{\rm max}$                      	& $3\times10^3$ & $5\times10^3$ \\      
$R_{\rm diss} [\rm parsec]$                     & 0.17 & 0.31  \\     
$R_{\rm BLR} [\rm parsec]$                      & 0.22 & 0.27 \\    
\tableline
$P_{\rm e} [\rm erg~s^{-1}$, in log scale$]$    & 44.94 & 45.13  \\      
$P_{\rm B} [\rm erg~s^{-1}$, in log scale$]$    & 45.38 & 45.77 \\      
$P_{\rm r} [\rm erg~s^{-1}$, in log scale$]$    & 45.70 & 46.37 \\      
$P_{\rm p} [\rm erg~s^{-1}$, in log scale$]$    & 47.45 & 47.12 \\      
$P_{\rm TOT} [\rm erg~s^{-1}$, in log scale$]$  & 47.47 & 47.14 \\
\tableline
\tableline
\end{tabular}
}
\captionof{table}{Table of comparison for the parameters used/derived from the SED modeling of J151002+570243 from \citet{2017ApJ...837L...5A} and this work.\label{tab:sed_par_nus_TXS}}
\end{minipage}
\hspace{0.2cm}
\begin{minipage}{0.48\textwidth}
\centering
\includegraphics[width=0.8\textwidth]{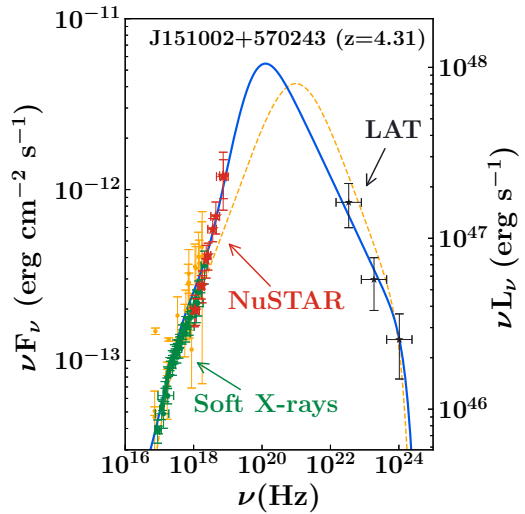}
\captionof{figure}{Zoomed high-energy SED of J151002+570243. The yellow/orange data points are the archival observations and the yellow/orange line is the SED model derived by \citealp{2017ApJ...837L...5A}. The blue line is the best fit model we obtained using combined soft X-ray (green), hard X-ray (red) and \gm-ray (black) data.\label{fig:sed_par_nus_TXS}}
\end{minipage}
\end{minipage}
\end{center}
\end{table*}

\section{Discussion}\label{sec:disc}
%%%%%%%%%%%%%%%%%%%%%%%%%%%%%%%%%%%%%%%%%%%%
High-redshift MeV blazars are among the most extreme sources of the blazar class. Harboring powerful relativistic jets and supermassive black holes at their center, they provide a unique opportunity to study jetted sources in the early Universe and are ideal to address open issues such as the accretion disk-jet connection and the growth and evolution of supermassive black holes at the dawn of the Universe.
Since in these sources the high-energy radiation dominates the bolometric luminosity, in order to understand the physics of the jets and the properties of the emitting leptonic population, a good coverage at these frequencies is required.
Awaiting an all sky MeV instrument (e.g.\ AMEGO, \citealp{1748-0221-12-11-C11024}, e-ASTROGAM, \citealp{2018JHEAp..19....1D}) to sample the peak of the IC emission, the most effective way to study the high-energy part of the SED is via a combination of X- and \gm-ray data.
As a result of the drift of the non-thermal SED peaks to lower frequencies, these high-redshift sources are brighter at X-ray frequencies, displaying a hard X-ray photon index ($\Gamma_{\rm X}\lesssim 1.5$) and a soft \gm-ray one ($\Gamma_{\gamma} \gtrsim2.3$).  
Four of the objects in our study already had \gm-ray data available \citep{2017ApJ...837L...5A} and, most importantly, we were able to obtain \nustar~observations for all six of them. All our sources show X-ray photon indices $<2$, as can be seen in Table~\ref{tab:results}, and soft \gm-ray ones ($\Gamma_{\gamma}>2.5$, \citealp{2017ApJ...837L...5A}). We point out that the sources J013126-100931 and J212912-153841 are detected in the BAT 105-months catalog \citep{2018ApJS..235....4O}, with a $\Gamma_{\rm 15-150\,keV}$ of $1.81^{+0.53}_{-0.48}$ and $1.79^{+0.48}_{-0.43}$, respectively. The unprecedented sensitivity of \nustar~allows us to accurately measure the hard X-ray photon index ($\Gamma_{\rm X}\sim1.55\pm0.1$ and $\Gamma_{\rm X}\sim1.56\pm0.03$, respectively, see Table~\ref{tab:results}).
Furthermore, \nustar~data are key to sample the rising part of the IC spectrum up to 60-70\,keV and, in combination with \fermi~ones, enable us to better constrain the peak location and consequently the bulk Lorentz factor and the shape of the electron energy distribution underlying this powerful emission. To highlight the importance of \nustar, in Figure \ref{fig:sed_par_nus}-\ref{fig:sed_par_nus_PKS} we show the zoomed SED for the three targets studied by \citet{2017ApJ...837L...5A}. In these plots, the yellow dotted line represents the EC model taken from their paper, while our EC model is shown as the blue solid line. Note that \citet{2017ApJ...837L...5A} performed the modeling using only X-ray archival observation (shown by yellow data points), whereas we benefited from the availability of \nustar~observations (red stars). 
SED parameters derived in this work and in \citet{2017ApJ...837L...5A} are also reported in Table~\ref{tab:sed_par_nus}-\ref{tab:sed_par_nus_PKS} for comparison. It is evident how \nustar~data enables us to sample the IC emission closer to the peak than previously possible, better constraining the slopes of the radiating leptonic population. For example, looking at J151002+570243 it can be seen how the value of $p$ drastically changes from 1.8 to 0.5, almost a factor of four difference from the previous modeling. These constraints translate into a more precise determination of the IC peak location and luminosity. We point out, as extreme examples, how our modeling predicts the IC peak of J151002+570243, or the IC peak luminosity of J064632+445116, to be nearly an order of magnitude lower than previously found only using soft X-ray data. The location of the emission regions is now more accurately constrained and found for all three sources to lie outside the BLR. These factors in turn produce a more accurate measure of the jet power and particle energy distribution. The substantial difference between the two EC models reflects the necessity of having hard X-ray coverage. 
\begin{table*}[t!]
\begin{center}
\begin{minipage}{\textwidth}

\begin{minipage}{0.5\textwidth}
\resizebox{\textwidth}{!}{
\begin{tabular}{lcc}
\tableline
\tableline
\multicolumn{3}{c}{J212912-153841} \\
\tableline
\tableline
Parameter & \citet{2017ApJ...837L...5A} & This work \\ 
\tableline                                                         
\tableline
$p$  						& 2.2 & 2.1 \\      
$q$  						& 4.5 & 4.0 \\      
$B$ [Gauss]                                     & 1.3 & 0.8 \\      
$U_{e}$erg cm$^{-3}$             		& 0.002 & 0.009 \\       
$\Gamma$                                   	& 14 & 10    \\      
$\gamma_{\rm min}$                      	& 1 & 4     \\      
$\gamma_{\rm break}$                      	& 51 & 60.32 \\      
$\gamma_{\rm max}$                      	& $2\times10^3$ & $3\times10^3$ \\      
$R_{\rm diss} [\rm parsec]$                     & 0.59 & 0.98  \\     
$R_{\rm BLR} [\rm parsec]$                      & 0.79 & 0.72 \\    
\tableline
$P_{\rm e} [\rm erg~s^{-1}$, in log scale$]$    & 45.14 & 45.89  \\      
$P_{\rm B} [\rm erg~s^{-1}$, in log scale$]$    & 46.58 & 46.33 \\      
$P_{\rm r} [\rm erg~s^{-1}$, in log scale$]$    & 46.31 & 48.11 \\      
$P_{\rm p} [\rm erg~s^{-1}$, in log scale$]$    & 47.89 & 48.12 \\      
$P_{\rm TOT} [\rm erg~s^{-1}$, in log scale$]$  & 47.92 & 47.99\\
\tableline
\tableline
\end{tabular}
}
\captionof{table}{Table of comparison for the parameters used/derived from the SED modeling of J212912-153841 from \citet{2017ApJ...837L...5A} and this work.\label{tab:sed_par_nus_PKS}}
\end{minipage}
\hspace{0.2cm}
\begin{minipage}{0.48\textwidth}
\centering
\includegraphics[width=0.8\textwidth]{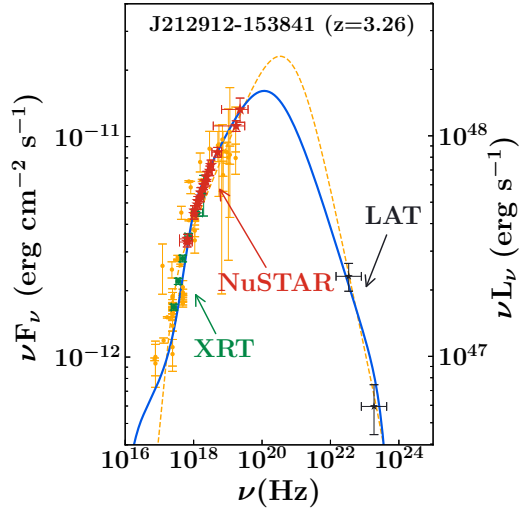}
\captionof{figure}{Zoomed high-energy SED of J212912-153841. The yellow/orange data points are the archival observations and the yellow/orange line is the SED model derived by \citealp{2017ApJ...837L...5A}. The blue line is the best fit model we obtained using combined soft X-ray (green), hard X-ray (red) and \gm-ray (black) data. \label{fig:sed_par_nus_PKS}}
\end{minipage}
\end{minipage}
\end{center}
\end{table*}

\nustar~has also proven to be essential in high-redshift blazar studies since, combined with soft X-ray observation, it has enabled us to detect peculiar X-ray features of these sources, such as spectral variability \citep[see][]{2015ApJ...807..167T, 2016MNRAS.462.1542S} and spectral flattening \citep[see][]{2016ApJ...825...74P}. 
The first property is important for various reasons. It could relate to intrinsic changes in the electron distribution (hence injected power) or could be key to unveil acceleration processes happening in the jet, such as magnetic reconnection or shocks \citep[e.g.,][]{2001MNRAS.325.1559S, 2011A&A...525A..40M, 2015arXiv150207882R, 2019MNRAS.482...65C}. Furthermore, it has been noticed that often X-ray variability correlates with \gm-ray one, indicating that the assumption of single electron population models are reliable. However, soft X-ray observations have sometimes not been reflective of this variation. 
On one hand, sensitivity limits of currently operating instruments could prohibit us from seeing such variability. On the other,
 the low-energy electron population emitting at these frequencies is not expected to vary significantly (see Section~\ref{sec:softvar}), while hard X-ray could show the signature of such variation \citep[see discussion in ][]{2016MNRAS.462.1542S}. Moreover as these sources become fainter in the \gm-rays, observing variability in this high-energy regime becomes a challenge \citep{2018ApJ...853..159L}. For all our sources, \nustar~data do not show any variability in the observation period. 
In the soft X-ray regime, we find both flux and photon index variability only in the case of J212912-153841 during the different observing epochs.  

\begin{figure}[t!]
\centering
\includegraphics[width=\columnwidth]{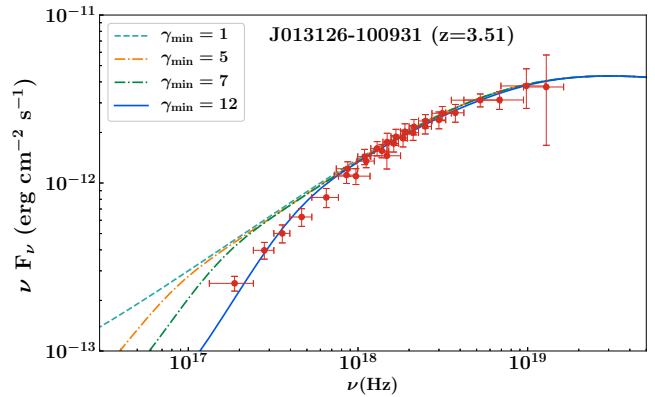}
	\caption{Zoomed SED of J013126$-$100931, showing the X-ray spectrum. The different lines represent the modeling done with various $\gamma_{\rm min}$ values (as labeled). As can be seen, good X-ray coverage allows us to constrain the value of $\gamma_{\rm min}$which, due to the break in the soft X-ray spectrum, is found to be $\sim12$. \label{fig:gamma_min}}
\end{figure}

The second property instead finds its cause in different scenarios. It could be related to the intrinsic behavior of the emitting electron population \citep[see][]{2016ApJ...825...74P} or could be attributed to an intrinsic absorber along the line of sight, such as the warm-hot IGM (see \citealp{2018A&A...616A.170A}), or a combination of the two.
The combination of soft X-ray and \nustar~data enables us to detect spectral curvature in the X-ray spectra of three of our sources (see Section~\ref{sec:xray_spec}). 
Within the available statistics, it appears that the softening below $\sim2$\,keV in J212912-153841 is a combination of both an intrinsic absorber probably located along the line of sight and a break in the electron distribution. The curvature below $\sim 5$\,keV for J013126-100931 instead is best explained by an intrinsic curvature of the leptonic distribution. Importantly, the shape of the X-ray spectrum constrains the behavior of the electrons underlying this emission. Indeed, if the X-rays are produced via EC rather than SSC, it is possible to constrain the electrons low-energy cut-off, as can be seen in Figure~\ref{fig:gamma_min}. In fact, for values of $\gamma_{\rm min}>1$, our model predicts a significant break which is seen in three of our sources. It follows that joint soft X-ray and \nustar~observations are crucial in determining the minimum energy of the leptonic population underlying the emission, which in turns regulates the jet power. 

J151002+570243, the farthest blazar ever detected in \gm-rays, is also the most peculiar of our sources in the X-ray regime. Indeed its X-ray spectrum not only favors a softening around $\rm\sim1\,keV$, but also shows a hardening $\rm>6\,keV$, in the energy range covered by \nustar, with extreme photon index of 0.94. 
We note that previous studies of this source in the soft X-ray regime found it to have a hard X-ray photon index ($\Gamma_{\rm X}=1.55\pm0.05$, \citealp{2013ApJ...763..109W}). This is another instance that reflects the necessity of \nustar~data to constrain the true shape of the X-ray emission, hence all jet parameters associated with it.

\begin{figure*}
\centering
\includegraphics[width=0.45\textwidth]{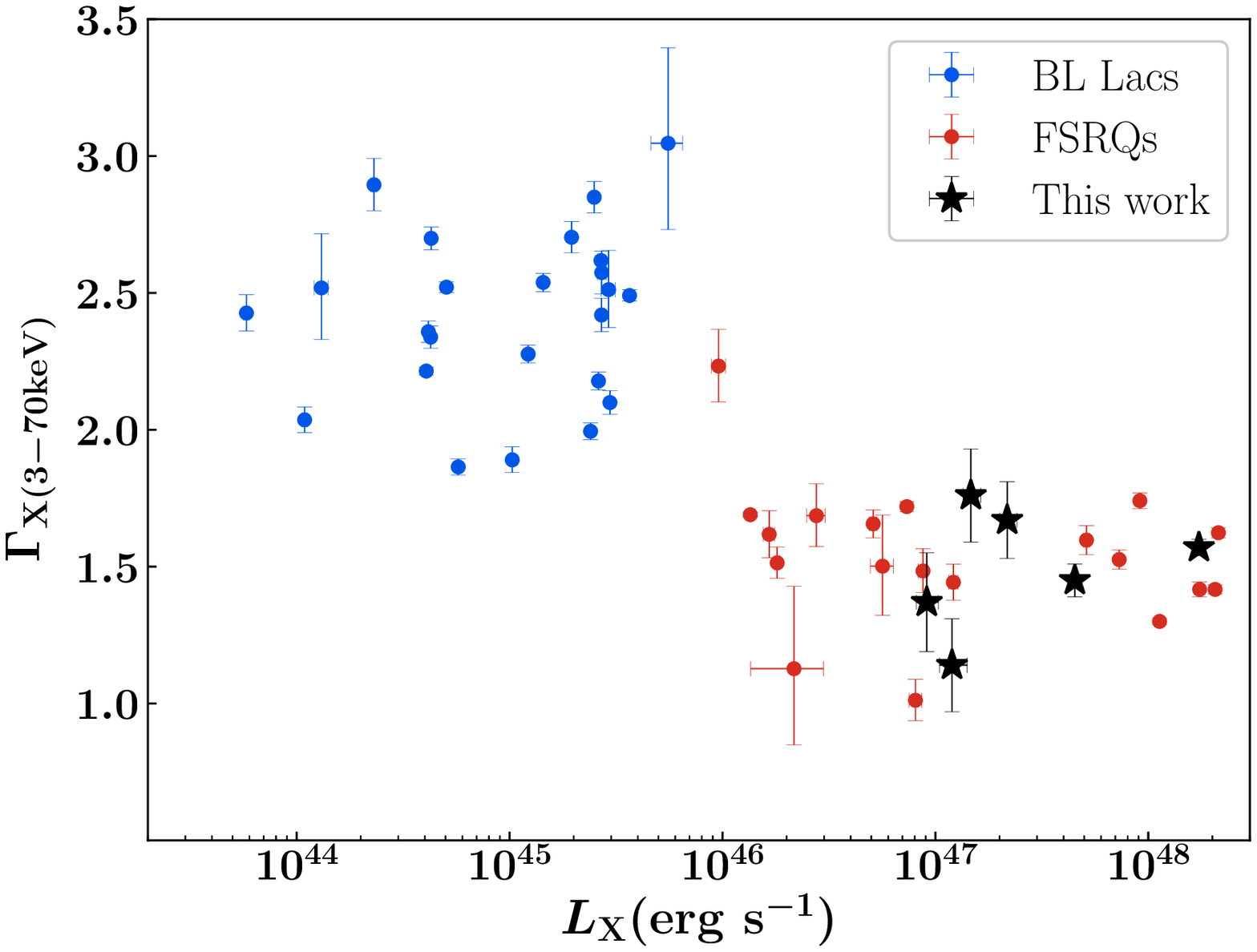}
\includegraphics[width=0.45\textwidth]{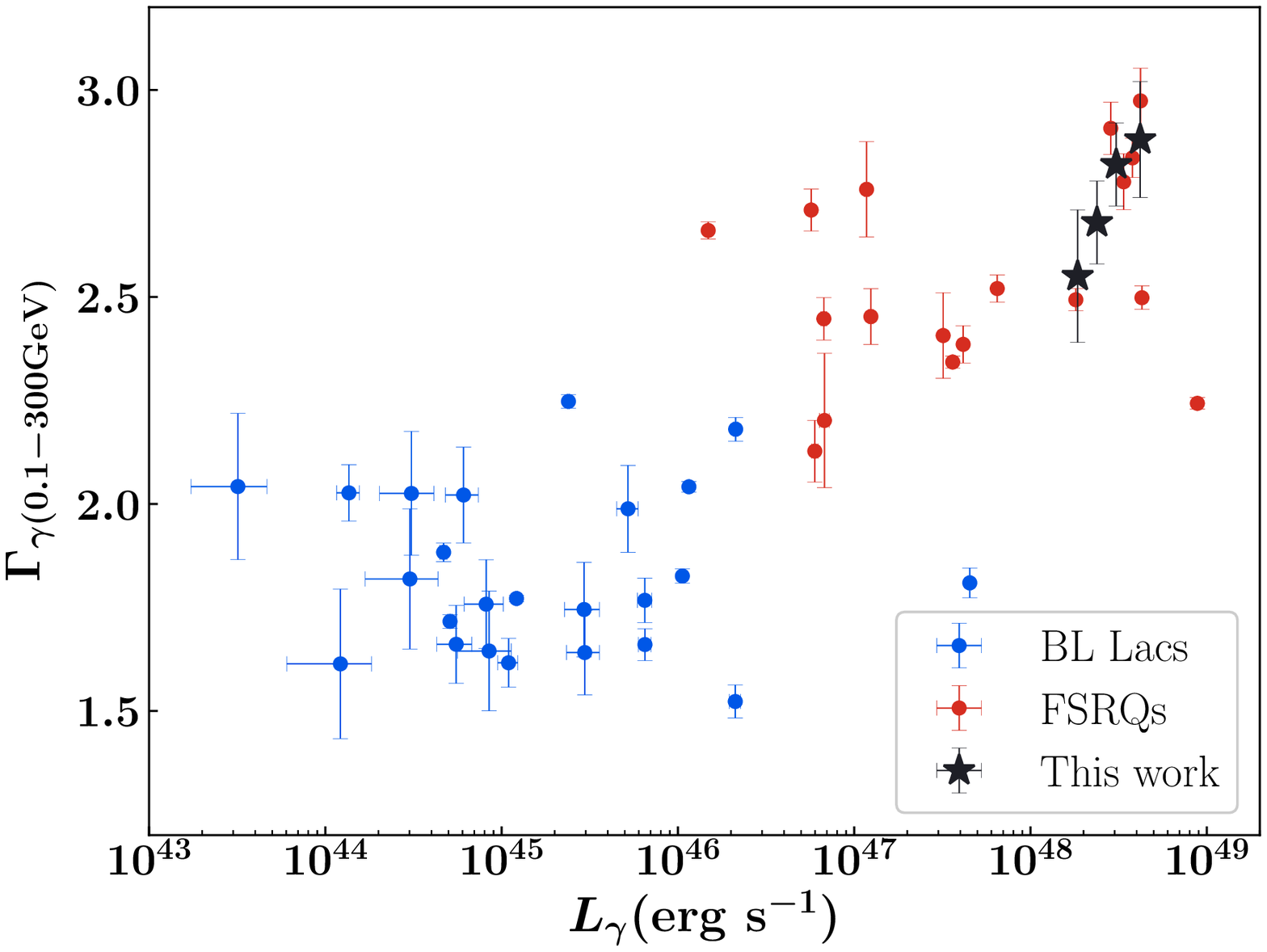}
	\caption{{\bf Left panel:} \nustar~luminosity versus X-ray spectral index for all BL Lacs and FSRQs listed in the 3LAC detected by \nustar~(Marcotulli et al., in preparation). Our sources fall in the lower right corner of the graph, having high X-luminosities and hard $\Gamma_{\rm X}$, therefore in agreement with the most powerful blazars. {\bf Right Panel:} \fermi-LAT luminosity versus \gm-ray spectral index for the same sources. As can be seen, our high-redshift blazars belong to the most powerful FSRQs, with high \gm-luminosities and very soft $\Gamma_{\gamma}$. \label{fig:lum_ind}}
\end{figure*}

From a modeling point of view, all of our sources have SEDs comparable to the most powerful LSP FSRQ blazars. They show the typical non-thermal humps, as well as the one produced by the accretion disk, unveiled by the SED frequency shift. The low-energy emission is therefore well interpreted by synchrotron process and peaks in the sub-mm regime. Thanks to the optical data obtained with SARA we are able to detect the peak of the disk emission and therefore to estimate disk luminosity and black hole masses from modeling\footnote{We note that our sources are assumed to be relatively efficient, even though remaining below the Eddington limit. We therefore impose that $10^{-2} L_{\rm Edd}< L_{\rm disk} < L_{\rm Edd}$.}.  
The high-energy part of the SED, covered by \swift-XRT, \nustar~and \fermi, instead peaks in (or close to) the MeV energy range and is explained by EC radiation. The X-ray spectra are very hard, which in turn suggests that EC process is dominant with respect to the SSC \citep[see][for a detailed discussion]{2016ApJ...826...76A}. In fact, the SSC is predicted by our modeling to be low for all of the sources, as can be seen in Figure~\ref{fig:sed}. 
In Figure \ref{fig:lum_ind}, the six sources here are compared to the \fermi-LAT blazars with available \nustar~observations (Marcotulli et al.\ in preparation).
Our sources fall in the same region as the most powerful FSRQs in both plots, showing high X-ray luminosities and hard X-ray indices as well as high \gm-ray luminosities and soft \gm-ray indices.
In all six of them, the IC peak luminosity dominates the synchrotron one (CD$>$1). In the adopted model, the radiative energy densities of the different AGN components are a function of the distance of the emission region from the central black hole. It follows that we can determine the location of the emission region. For our sources we find it resides outside the BLR and inside the torus (thus in agreement with \citealp{2018MNRAS.477.4749C}). Therefore the torus is the predominant source of radiation for the EC. 
Overall, our sources resemble the most extreme MeV blazars. 

It has been found that, for \gm-ray detected blazars up to $z=3$, the accretion disk luminosity is lower than the jet power \citep{2014Natur.515..376G}. More recent results show that a positive correlation remains for \gm-quiet sources (i.e.\, blazars not detected in \gm-rays) beyond $z=3$ \citep{2017ApJ...851...33P}. However, as discussed above, having \gm-ray data is crucial to accurately determine the jet power. 
Therefore, with good \gm- and X-ray data in hand, as well as optical ones, we can insert our sources in the context of the disk-jet relation. In Table~\ref{tab:sed_par} we see how all six have jet powers exceeding disk luminosities. This further hints to the idea that there has to be an extra reservoir of energy powering the jet rather than the accretion disk alone (e.g.\ could be supplied by the spinning black hole, \citealp{1977MNRAS.179..433B}). However, we need to be cautious and take into account uncertainties in our measurements. For example, in our treatment we do not consider electron positron pairs, which, if present, would reduce the estimated jet power by a factor $n_e/n_p$ \citep[$n_e=n_{e^+}+n_{e^-}$, see][]{doi:10.1093/mnras/stw2960}. Nevertheless, to avoid the Compton Rocket effect that would otherwise break the jet \citep[see][]{2010MNRAS.409L..79G}, the number of pairs could not outnumber by a large amount the number of protons. Moreover, another effect that could reduce the jet power is if we consider it to have a spine-sheath structure \citep{2016MNRAS.457.1352S}. An underestimation of the accretion disk luminosity could also lead to a false positive relation. So these caveats have to be kept in mind when analyzing the relation between jet power and disk luminosity in blazars.

An important jet parameter that is obtained via modeling is the bulk Lorentz factor. In fact, assuming that the jet points towards the observer at an angle $\lesssim 1/\Gamma$, by a simple geometrical argument, we can estimate the size of the parent population of these sources. It turns out that the detection of one jetted source implies the existence of $2\Gamma^2$ similar sources, at the same $z$, with similar black hole mass, but with jets pointing away from the observer. From various studies, it has been found that high-redshift blazars host supermassive black holes. Using the $2\Gamma^2$ correction, \citealp{2015MNRAS.446.2483S} derived the comoving number density of billion solar mass black holes and found that for radio-quiet blazars (i.e.\ that host weak or no jets) it peaks at $z\sim2$, while for radio-loud blazars (i.e.\ that have strong relativistic jets) it peaks at $z\sim4$. 
This fact challenges our understanding of black hole growth and evolution in the early Universe since, as pointed out by \citealp{2015JHEAp...7..163G},  
simple accretion (even considering highly massive seeds at $z=20$) cannot account for $M_{\rm BH}>10^9\Msun$ at $z>4$. On account of the fact that the number density strongly depends on the number of sources detected per redshift bin, finding more such sources is crucial to understand the physics of early black hole accretion. 
The apparent dichotomy in \citealp{2015MNRAS.446.2483S} between radio-quiet and radio-loud sources further hints 
to a connection between rapid black hole growth and presence of powerful jets.
From SED modeling, five of our sources have $M_{\rm BH}>10^9\Msun$\footnote{We consider the masses derived by SED modeling for the sources for which we have obtained optical data.}. We use the prescription in \citealp{2017ApJ...837L...5A} to estimate the number density ($n$) of these $M_{\rm BH}>10^9\Msun$ radio-loud blazars in the redshift bin $z=[3,4]$\footnote{Since four of these sources are the ones discovered by \citealp{2017ApJ...837L...5A}, we use the same parameters listed in the paper.}.
This number was derived by \citealp{2015MNRAS.446.2483S} to be 50\,Gpc$^{-3}$ and \citet{2017ApJ...837L...5A} increased this estimate by 18\,Gpc$^{-3}$. From our calculation, we find $n\sim37 \rm\,Gpc^{-3}$ which leads the total space density value to be 87 Gpc$^{-3}$. This seems to further confirm a connection between radio-loud phase and rapid accretion. Only detecting more sources would improve our understanding of this rapid and early black hole accretion.

\section{Conclusions}
In this work, the multiwavelength analysis of six high-redshift ($z>3$) blazars, four of which were recently discovered to be \gm-ray loud by \citep{2017ApJ...837L...5A}, is presented. Employing the excellent capabilities of \nustar, we are able to precisely study X-ray properties of these sources. In combination with quasi-simultaneous \swift~and \fermi-LAT observations, we accurately constrained their jet properties (i.e. jet power, underlying electron distribution, location of the emission region). Simultaneous optical monitoring with SARA reveals the peak of the disk emission through which we can determine the accretion disk luminosities and black hole masses. 
The main results are summarized here:
\begin{itemize}
\item \nustar~spectra are very hard ($\Gamma\lesssim1.5$) and do not show variability. Combining \nustar~and sensitive soft X-ray observations, we detect spectral flattening in the X-ray spectra of three sources. The curvature in J013126-100931 can be explained by a property of the underlying electron distribution (i.e.\ depends on the minimum energy of the electron population). In the case of J212912-153841, it is attributed to a combination of both intrinsic curvature in the leptonic distribution and an intrinsic absorber in the IGM. The peculiar case of J151002+570243 ($z=4.31$) reveals two breaks in its X-ray spectrum, the first possibly attributed to the minimum energy of the electrons producing this emission and/or an IGM absorption, while the second can be explained in the context of the one-zone leptonic emission model.
\item Using a one-zone leptonic emission model, \nustar~and \fermi-LAT spectra allow us to precisely pinpoint the position of the high-energy peak, which is well described by EC emission of the relativistic electrons in the jet. This enables us to determine jet parameters such as the jet power and the emission region location (for all sources outside the BLR), as well as constrain the shape of the underlying electron distribution. 
\item Inserting our sources in the context of accretion disk connection, we find that all sources display jet powers larger than accretion disk luminosity, validating the possibility of extra source of power to the jet, such as the spinning black hole.
\item The optical data unveil the disk emission and enable us to estimate the disk luminosity and black hole mass. All sources have $\rm M>10^9\Msun$, which further raises the space density of supermassive black-holes in the redshift bin $z=[3,4]$ to $87\,\rm Mpc^{-3}$.
 
\end{itemize}

\acknowledgements
L.M., V.P. and M.A. acknowledge funding under NASA contract: NNX17AC35G.

The \textit{Fermi} LAT Collaboration acknowledges generous ongoing support
from a number of agencies and institutes that have supported both the
development and the operation of the LAT as well as scientific data analysis.
These include the National Aeronautics and Space Administration and the
Department of Energy in the United States, the Commissariat \`a l'Energie Atomique
and the Centre National de la Recherche Scientifique / Institut National de Physique
Nucl\'eaire et de Physique des Particules in France, the Agenzia Spaziale Italiana
and the Istituto Nazionale di Fisica Nucleare in Italy, the Ministry of Education,
Culture, Sports, Science and Technology (MEXT), High Energy Accelerator Research
Organization (KEK) and Japan Aerospace Exploration Agency (JAXA) in Japan, and
the K.~A.~Wallenberg Foundation, the Swedish Research Council and the
Swedish National Space Board in Sweden.
 
Additional support for science analysis during the operations phase is gratefully
acknowledged from the Istituto Nazionale di Astrofisica in Italy and the Centre
National d'\'Etudes Spatiales in France. This work performed in part under DOE
Contract DE-AC02-76SF00515.

We thank the \swift~team and the \swift~PI (B. Cenko) for promptly scheduling and executing the observations.

Reported work is in part based on observations obtained with the SARA Observatory 1.0 m, 0.96 m and 0.6 m telescopes respectively located at la Roque de los Muchachos Observatory, Canary Islands (SARA-ORM), at Kitt Peak National Observatory, Arizona (SARA-KP) and at Cerro-Tololo, Chile (SARA-CT), which are owned and operated by the Southeastern Association for Research in Astronomy (saraobservatory.org).

Part of this work is based on archival data, software or on–line services provided by the
ASI Data Center (ASDC). This research has made use of the XRT Data Analysis Software
(XRTDAS). 

\appendix 
\section{X-ray spectral fits}{\label{app:A}}

This Appendix contains the X-ray spectra of the six blazars as well as the residual of the fits in the same order as listed in Table~\ref{tab:results}. The green dotted line is the best fit found analysing the residuals as well as performing F-tests (see Section~\ref{sec:xray_spec}). 

\begin{figure}[h!]
\centering
\includegraphics[width=0.45\textwidth]{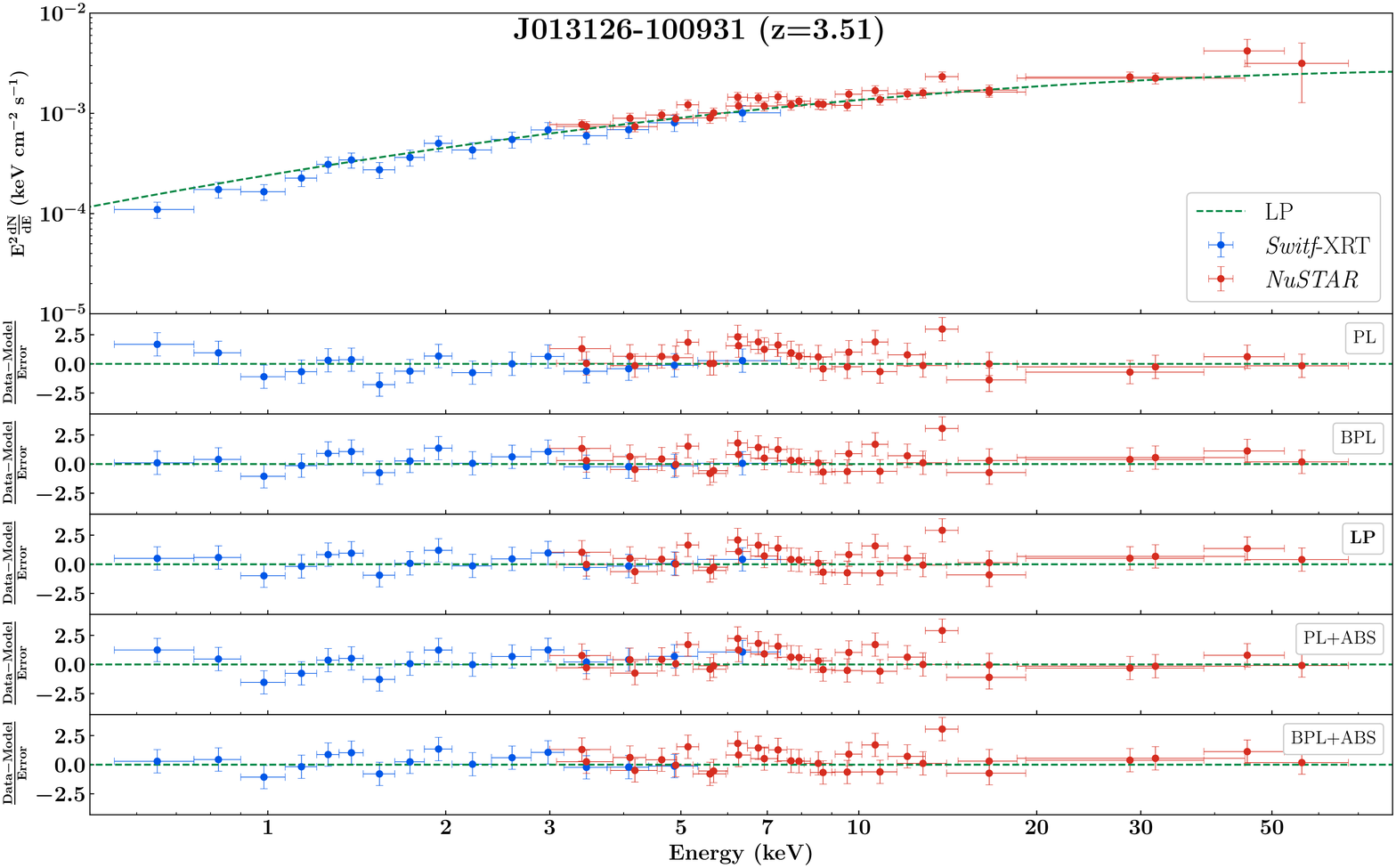}
\includegraphics[width=0.45\textwidth]{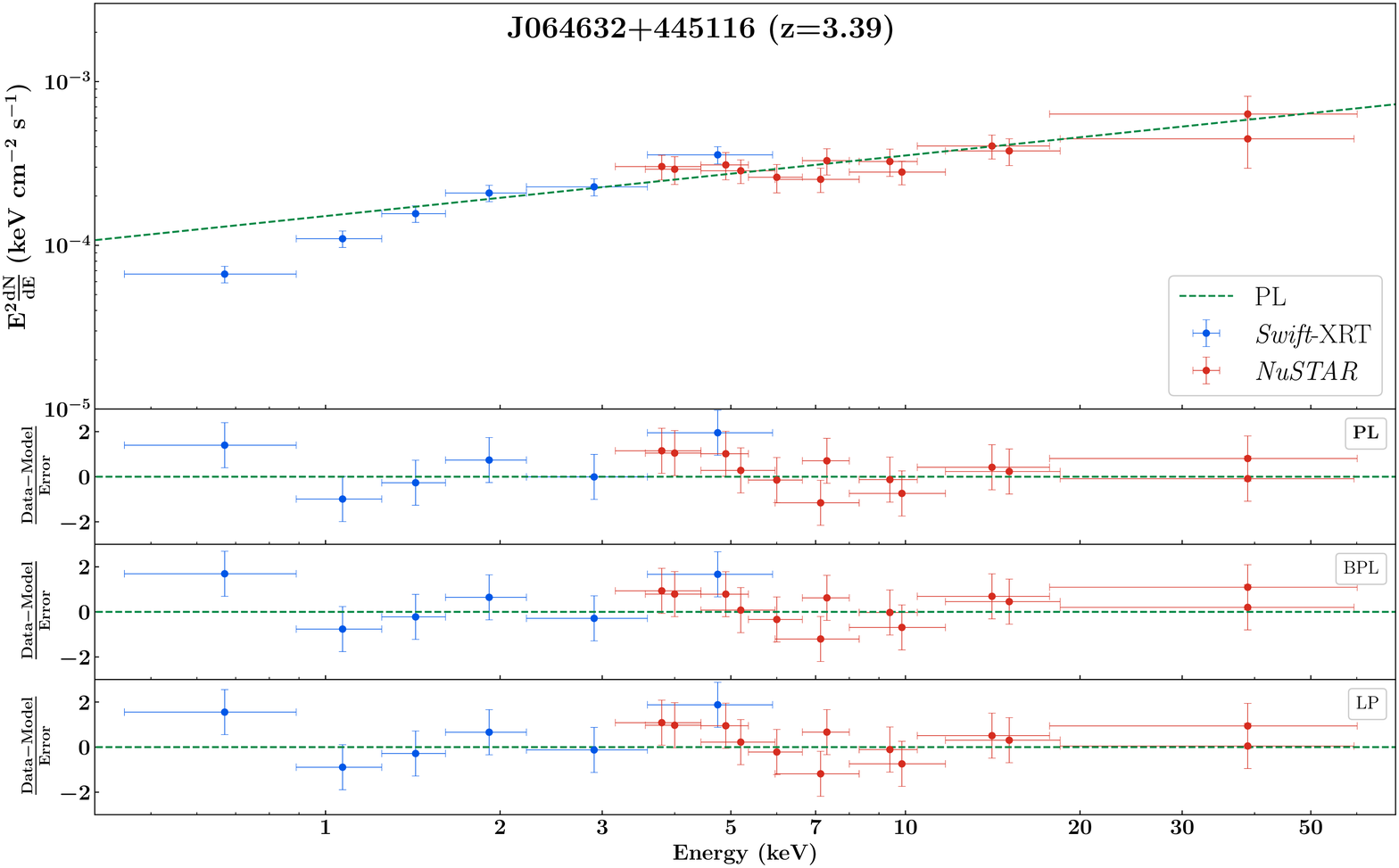}
\includegraphics[width=0.45\textwidth]{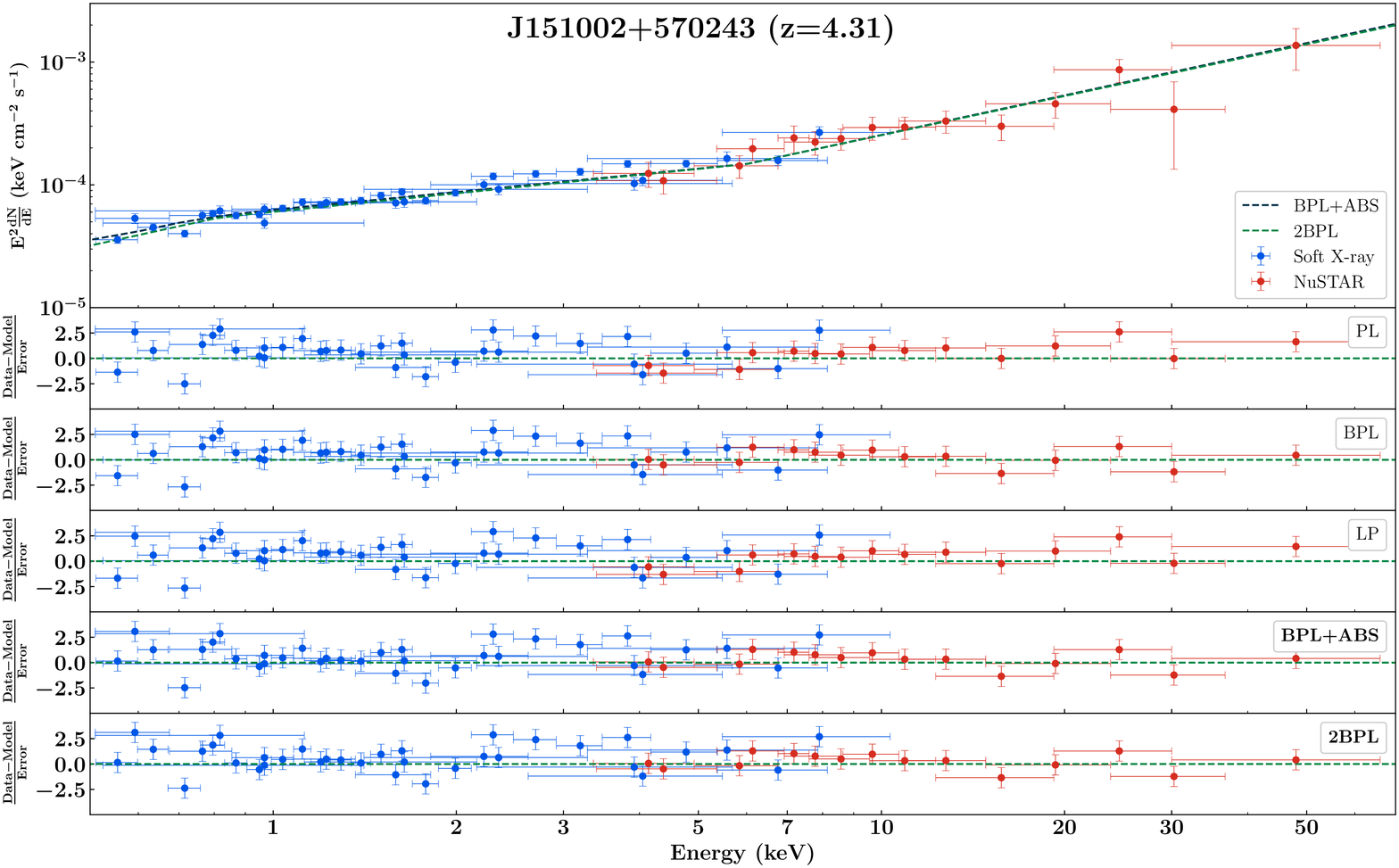}
\includegraphics[width=0.45\textwidth]{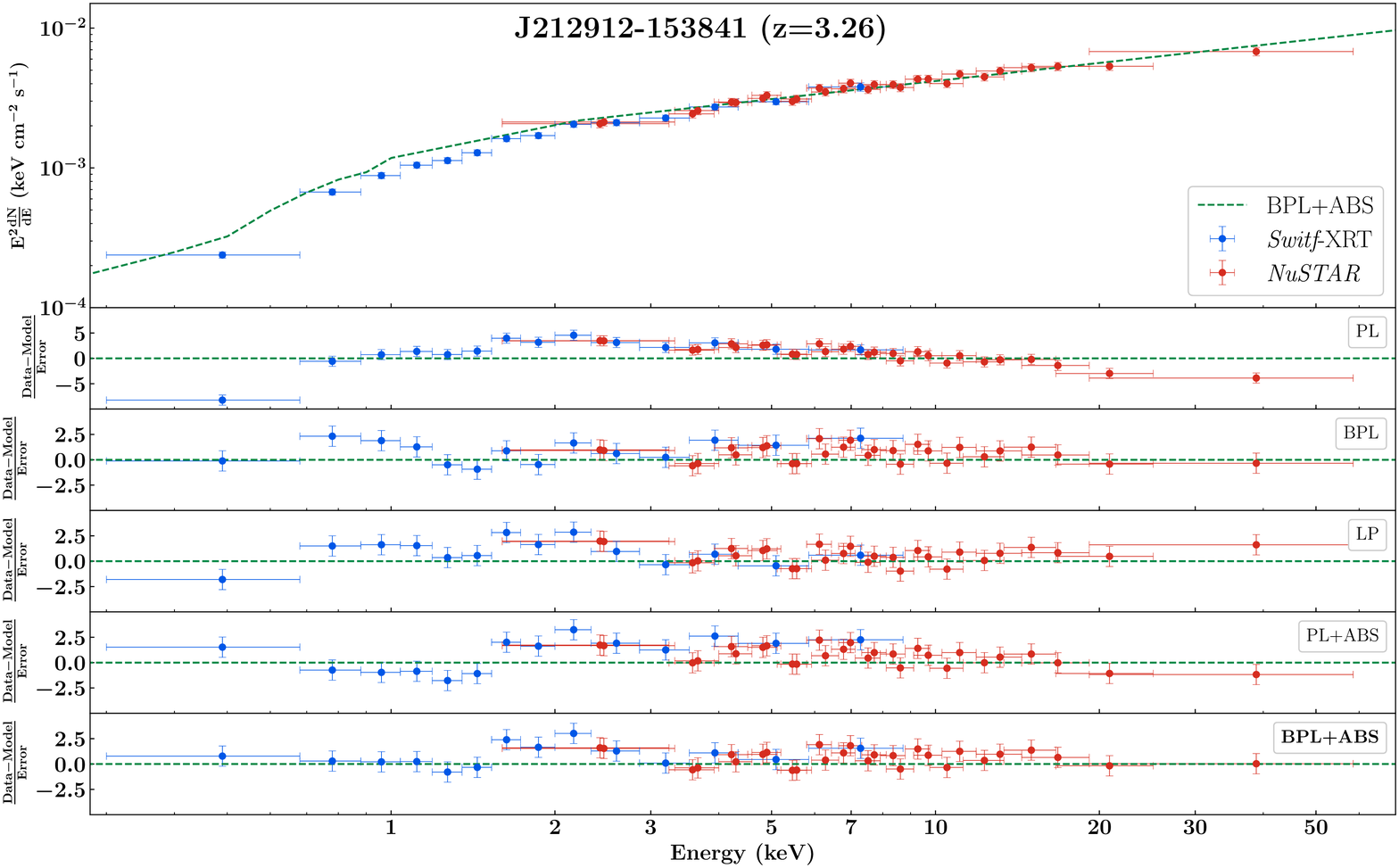}
\includegraphics[width=0.45\textwidth]{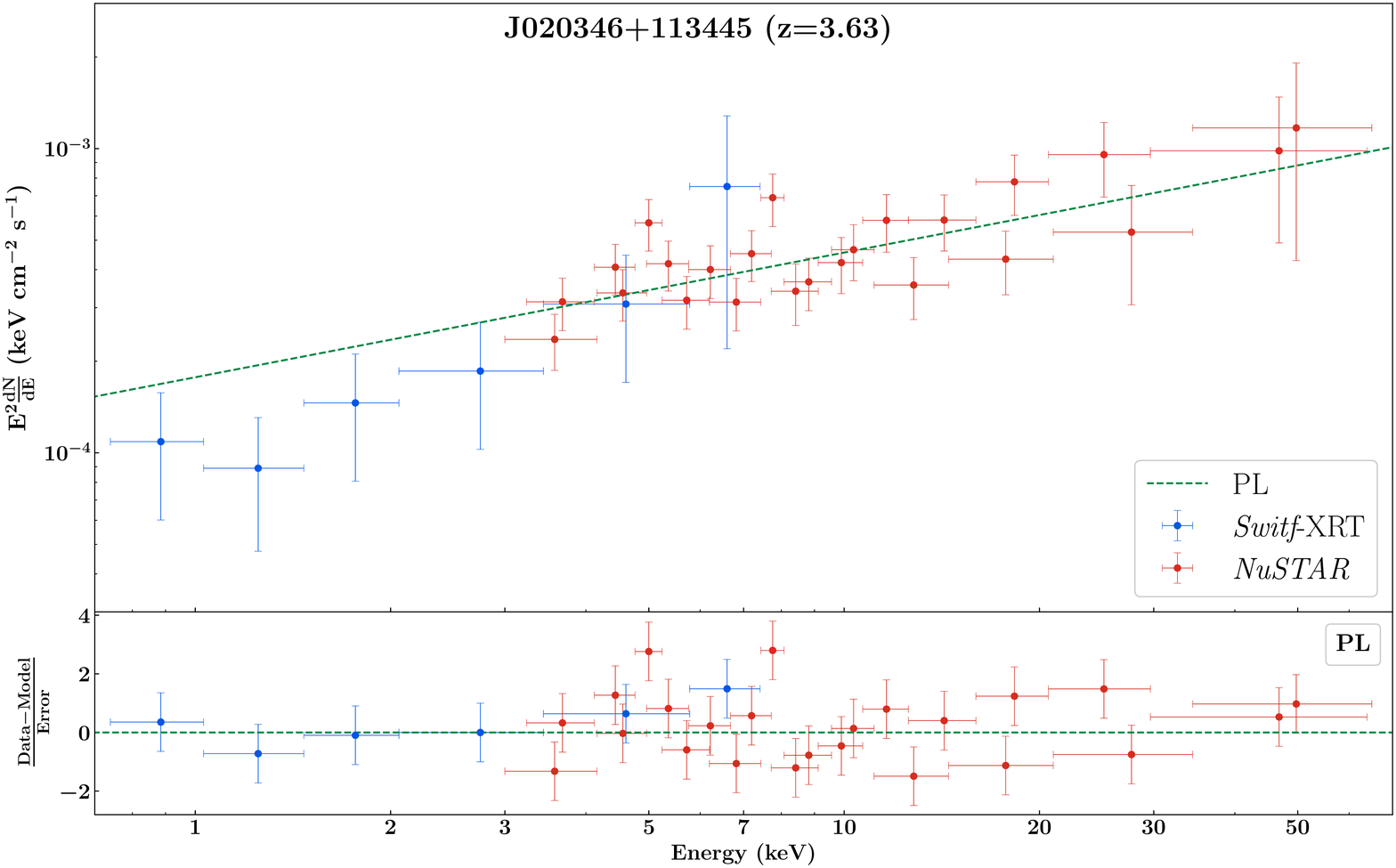}
\includegraphics[width=0.45\textwidth]{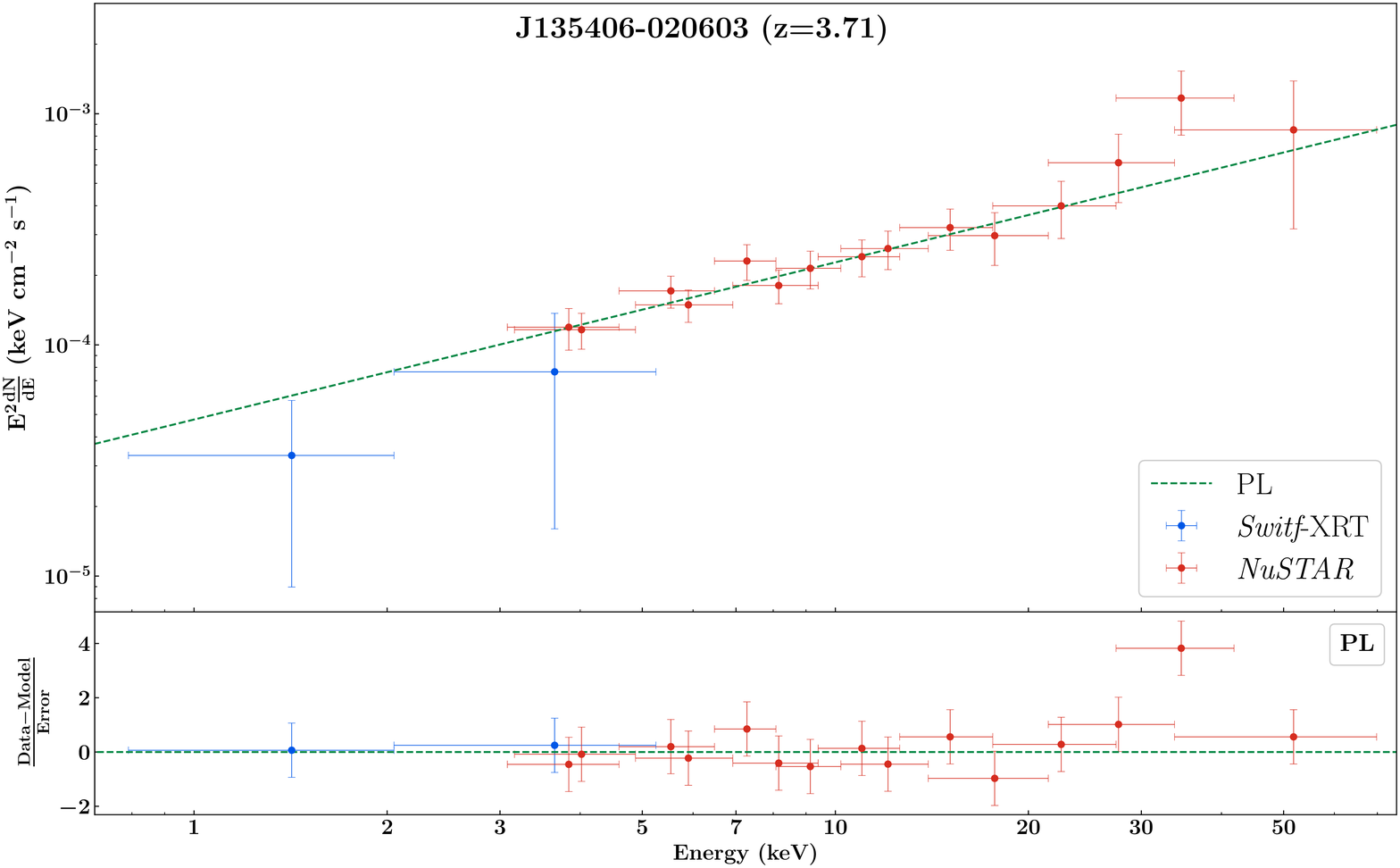}
\end{figure}

\newpage
\bibliographystyle{aasjournal}
\bibliography{MeV}

\end{document}